\documentclass[12pt]{iopart}

\usepackage{graphicx}
\usepackage{amsfonts}
\usepackage{amssymb}
\usepackage{url}
\usepackage{epstopdf}
\usepackage{color}
\usepackage{bm}
\usepackage[colorlinks=true,linkcolor=blue,citecolor=red]{hyperref}%
\usepackage{cite}
\usepackage{dsfont}

\def\L{\mathcal L}
\def\N{\mathcal N}

\def\n{\bm{n}}
\def\s{\bm{s}}
\def\x{\bm{x}}
\def\X{\bm{X}}
\def\W{\bm{W}}

\def\J{\bm{J}}

\def\pa{\partial\Omega}

\def\E{{\mathbb E}}

\def\P{{\mathbb P}}
\def\R{{\mathbb R}}
\def\Z{{\mathbb Z}}
\def\T{{\mathcal T}}
\def\L{{\mathcal L}}

\def\AA{\mathcal A}
\def\RR{\mathcal R}

\def\mmu{\bm{\mu}}

\def\M{{\mathcal M}}

\def\erf{\mathrm{erf}}
\def\erfc{\mathrm{erfc}}
\def\erfcx{\mathrm{erfcx}}
\def\ctanh{\mathrm{ctanh}}
\def\ctan{\mathrm{ctan}}
\def\tanh{\mathrm{tanh}}

\begin{document}

\title[An encounter-based approach for restricted diffusion...]{An encounter-based approach for restricted diffusion with a gradient drift}

\author{Denis~S.~Grebenkov}
 \ead{denis.grebenkov@polytechnique.edu}
% \email{denis.grebenkov@polytechnique.edu}
%\affiliation{
\address{
$^1$ Laboratoire de Physique de la Mati\`{e}re Condens\'{e}e (UMR 7643), \\ 
CNRS -- Ecole Polytechnique, IP Paris, 91120 Palaiseau, France}

\address{
$^2$ Institute for Physics and Astronomy, University of Potsdam, 14476
Potsdam-Golm, Germany}

\date{\today}

\begin{abstract}
We develop an encounter-based approach for describing restricted
diffusion with a gradient drift towards a partially reactive boundary.
For this purpose, we introduce an extension of the
Dirichlet-to-Neumann operator and use its eigenbasis to derive a
spectral decomposition for the full propagator, i.e., the joint
probability density function for the particle position and its
boundary local time.  This is the central quantity that determines
various characteristics of diffusion-influenced reactions such as
conventional propagators, survival probability, first-passage time
distribution, boundary local time distribution, and reaction rate.  As
an illustration, we investigate the impact of a constant drift onto
the boundary local time for restricted diffusion on an interval.  More
generally, this approach accesses how external forces may influence
the statistics of encounters of a diffusing particle with the reactive
boundary.
\end{abstract}


%02.50.-r       (Probability theory, stochastic processes, and statistics)
%05.40.-a 	Fluctuation phenomena, random processes, noise, and Brownian motion
%02.70.Rr       (General statistical methods)
%05.10.Gg 	Stochastic analysis methods (Fokker-Planck, Langevin, etc.) 

%02.50.Ey 	Stochastic processes  (Probability theory, stochastic processes, and statistics)


%\keywords{Boundary local time, Reflected Brownian motion, Diffusion-influenced reactions, 
%Surface reactivity, Robin boundary condition, Heterogeneous catalysis}

\submitto{\JPA}

\maketitle

\section{Introduction}
\label{sec:intro}

Many transport processes in nature and industry are described by an
overdamped Langevin equation for the random position $\X_t$ of a
particle at time $t$ or, equivalently, by the associated Fokker-Planck
equation for the probability density of that position
\cite{Carslaw,Risken,Gardiner,Redner,Schuss,Metzler,Lindenberg}.
In the presence of restricting boundaries or hindering obstacles, the
above descriptions have to be adapted to account for interactions of
the diffusing particle with that boundaries.  In physics literature,
one usually deals directly with the Fokker-Planck equation by imposing
appropriate boundary conditions, while its stochastic counter-part is
generally ignored.  At the same time, the stochastic differential
equation in a confining domain naturally yields additional information
on the statistics of encounters of the diffusing particle with the
boundary, the so-called boundary local time $\ell_t$.  In a recent
work \cite{Grebenkov20}, we investigated ordinary restricted diffusion
and showed numerous advantages of using the joint probability density
for the pair $(\X_t,\ell_t)$ to build up a new encounter-based
approach to diffusion-influenced reactions and other
diffusion-mediated surface phenomena.  In particular, surface
reactions can be incorporated explicitly via an appropriate stopping
condition for the boundary local time.  In this way, the bulk
dynamics, determined by the pair $(\X_t,\ell_t)$ in a confining domain
with a fully inert reflecting boundary, is disentangled from surface
reactions, which are imposed later on.  Moreover, this approach allows
one to implement new surface reaction mechanisms, far beyond those
described by the conventional Robin boundary condition.  For instance,
one can introduce an encounter-dependent reactivity, in analogy with
time-dependent diffusion coefficient for bulk dynamics.

In this paper, we extend the encounter-based approach proposed in
\cite{Grebenkov20} to more general diffusion processes with a gradient
drift.  Section \ref{sec:general} starts by recalling two conventional
descriptions of restricted diffusion and discussing the role of the
full propagator in the case of ordinary restricted diffusion.  After
this reminder, we present the main results by introducing an extension
of the Dirichlet-to-Neumann operator and using its eigenbasis for
deriving a new spectral decomposition for the full propagator, as well
as for various characteristics of diffusion-influenced reactions.  In
Sec. \ref{sec:interval}, we illustrate our general results in a simple
setting of restricted diffusion with a constant drift on an interval
(or, equivalently, between parallel planes).  The geometric simplicity
of this setting allows us to avoid technical issues and to get fully
explicit formulas that shed light onto the role of the drift onto
boundary encounters.  Section \ref{sec:discussion} presents a critical
discussion of the proposed approach and highlights its advantages,
drawbacks and limitations, as well as further perspectives.

\section{General spectral approach}
\label{sec:general}

\subsection{Two conventional descriptions}
 
\begin{figure}
\begin{center}
\includegraphics[width=80mm]{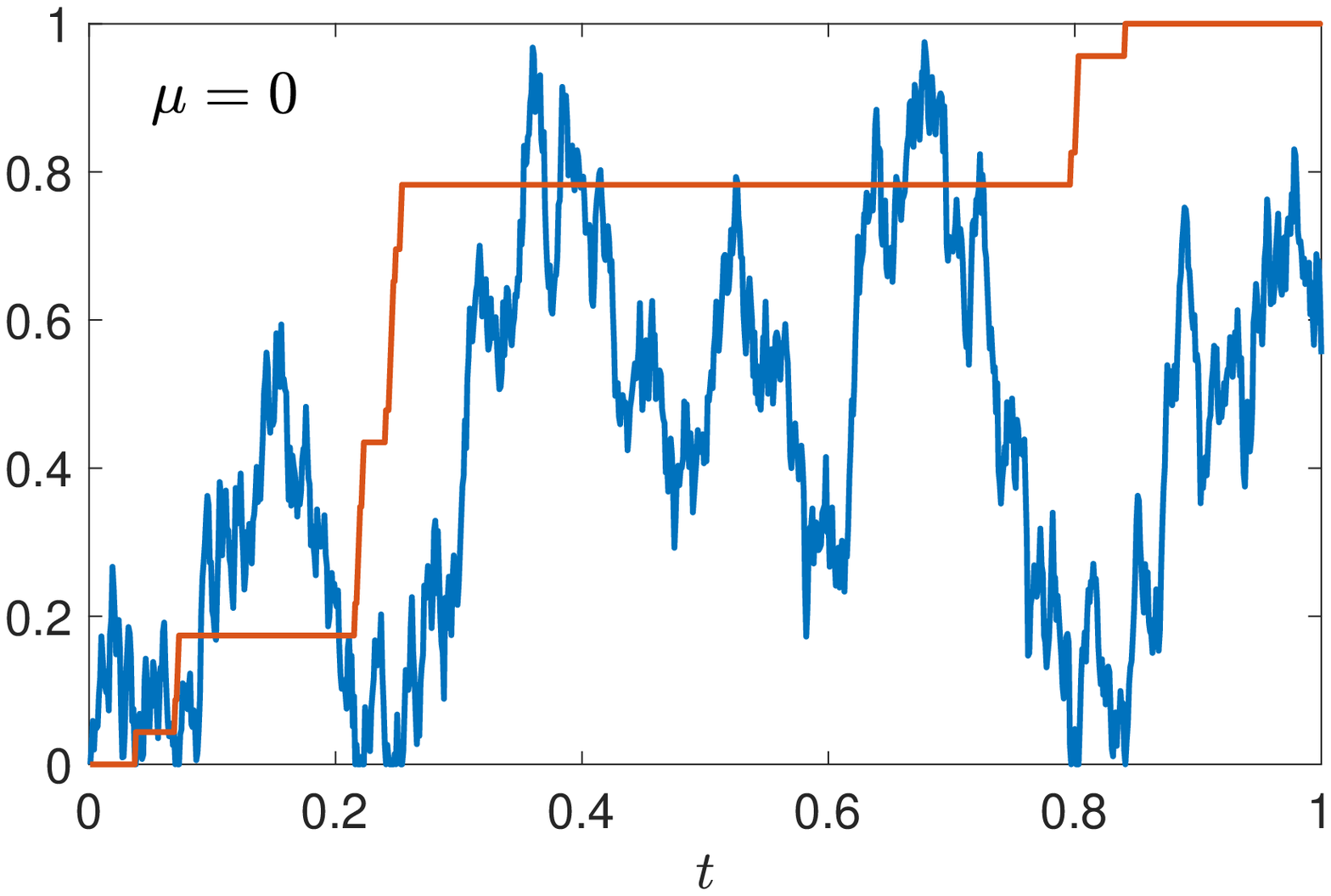} % simu_mu0.eps}
\includegraphics[width=80mm]{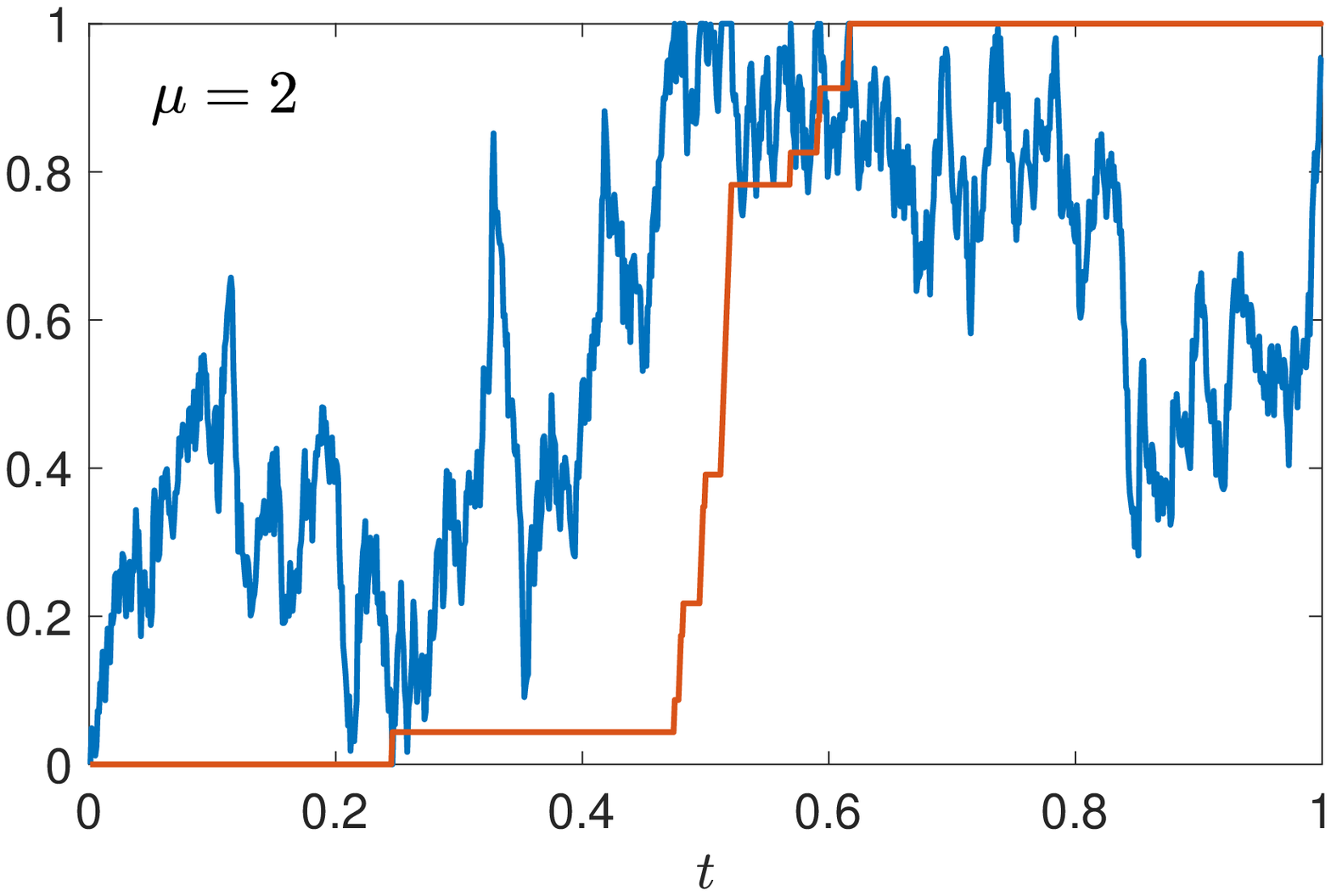} % simu_mu2.eps}
\includegraphics[width=80mm]{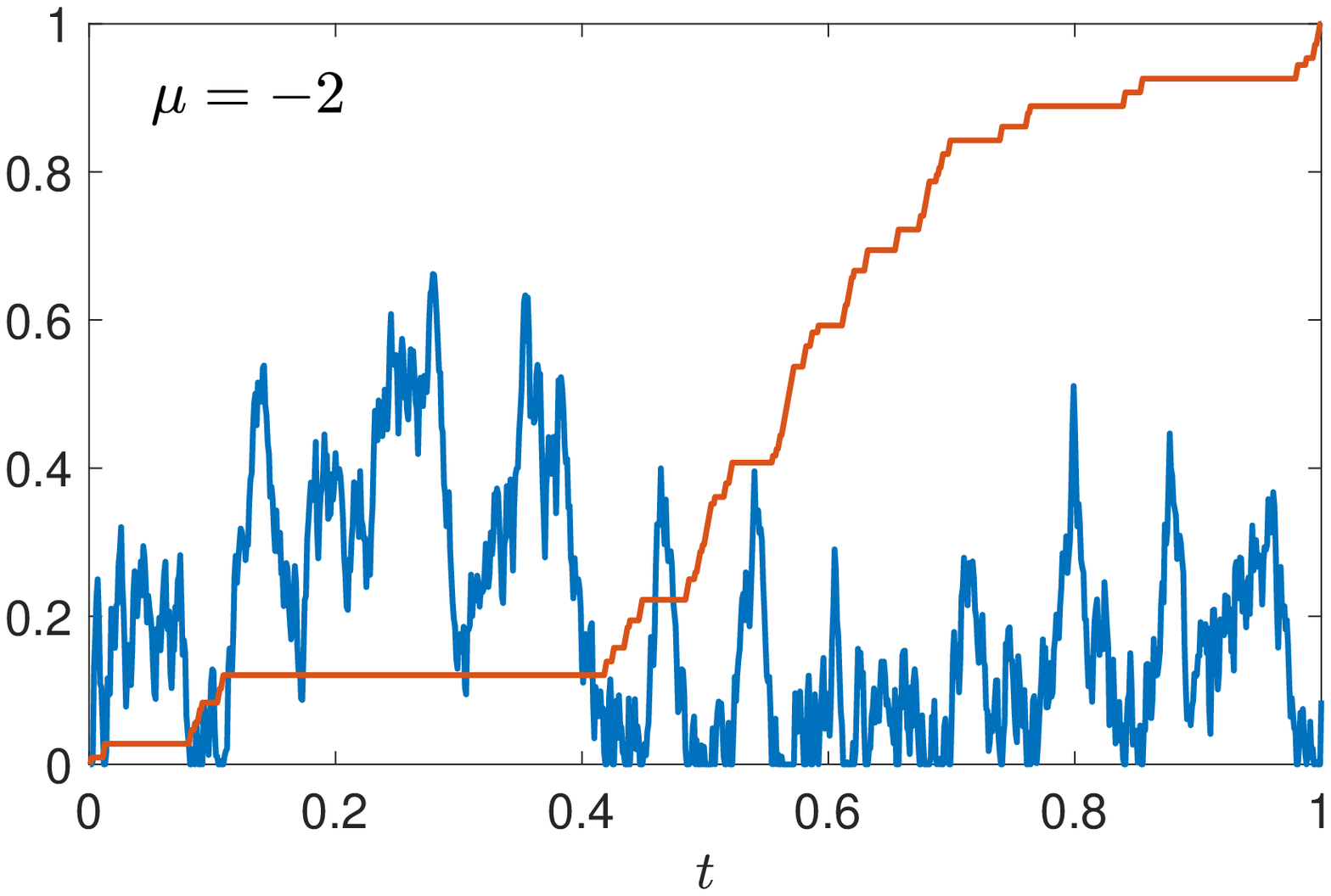} % simu_mu_2.eps}
\end{center}
\caption{
Simulated trajectory $X_t$ (blue line) of restricted diffusion on the
unit interval $(0,1)$ with the starting point $x_0 = 0$ and $D = 1$,
and the boundary local time $\ell_t$ (red line), rescaled by its value
$\ell_T$ at $T = 1$, for three values of the constant drift: $\mu = 0$
(top), $\mu = 2$ (middle) and $\mu = -2$ (bottom).  Each trajectory
was generated by Monte Carlo simulations with a time step $10^{-3}$. }
\label{fig:simu}
% [x,ell] = A_localtime8_drift_simu(x0,mu);
\end{figure}

We first recall two conventional descriptions of restricted diffusion
in a bounded domain $\Omega\subset\R^d$ with a smooth boundary $\pa$.
On one hand, it can be described by the Skorokhod stochastic
differential equation for the random position $\X_t =
(X_t^1,\ldots,X_t^d)$ of a diffusing particle at time $t$:
\begin{equation}  \label{eq:Skorokhod}
d\X_t = \mmu(\X_t) dt + \sqrt{2D}\, d\W_t - \n(\X_t) d\ell_t, \qquad \X_0 = \x_0 ,
\end{equation}
where $\x_0$ is the starting point \cite{Levy,Ito,Freidlin}.  Here the
infinitesimal displacement $d\X_t$ of the particle has three
contributions: (i) the deterministic part with a drift $\mmu(\x)$,
(ii) random (thermal) fluctuations with the standard Gaussian noises
$d\W_t$ whose amplitudes are determined by the diffusion coefficient
$D$, and (iii) reflections from the boundary $\pa$ along the unit
normal vector $-\n(\x)$ (oriented inward the domain), whenever the
particle hits the boundary.  Intuitively, the last term can be
understood as an infinitely strong but infinitely short-ranged force
that repels the particle from the boundary back to the bulk (see a
physical rational behind this force in SM.I of
\cite{Grebenkov20}).
Such instant reflections are governed by the boundary local time
$\ell_t$ -- a non-decreasing stochastic process (with $\ell_0 = 0$)
that increases at each encounter with the boundary (see
Fig. \ref{fig:simu}).  Curiously, the single Skorokhod equation
determines simultaneously two tightly related stochastic processes:
the position $\X_t$ and the boundary local time $\ell_t$.   Once
$\X_t$ is constructed, the boundary local time $\ell_t$ can be
obtained by rescaling the residence time of the particle in the thin
boundary layer of width $a$ (see
\cite{Darling57,Ray63,Knight63,Agmon84,Berezhkovskii98,Dhar99,Yuste01,Godreche01,Majumdar02,Benichou03,Condamin05,Condamin07,Burov07,Burov11}
and references therein):
\begin{equation}  \label{eq:ellt_def}
\ell_t = \lim\limits_{a\to 0} \frac{D}{a} \underbrace{\int\limits_0^t dt' \, \Theta(a - |\X_{t'} - \pa|)}_{\textrm{residence time}} ,
\end{equation}
where $\Theta(z)$ is the Heaviside step function ($\Theta(z) = 1$ for
$z > 0$ and $0$ otherwise), and $|\x - \pa|$ denotes the distance
between a bulk point $\x$ and the boundary $\pa$.  From
Eq. (\ref{eq:ellt_def}), one can see that the boundary local time
$\ell_t$ is indeed a non-decreasing process that remains constant when
$\X_t$ is the bulk, and increases only when $\X_t$ hits the boundary.
Even though $\ell_t$ has units of length, we keep using the canonical
term ``boundary local time''.  Note that $\ell_t/D$ has units of time
per length so that its multiplication by the width $a$ of a thin
boundary layer approximates the fraction of {\it time} that the
particle spent in that layer up to time $t$.

Throughout this work, we focus on the common physical setting when
the drift represents an external conservative force $\bm{F}(\x)$,
which can be written as the gradient of a potential $V(\x)$:
\begin{equation}
\mmu(\x) = \frac{\bm{F}(\x)}{\gamma} = \frac{-\nabla V(\x)}{\gamma} = - D \nabla \Phi(\x) ,
\end{equation}
where $\Phi(\x) = V(\x)/(k_B T)$ is the dimensionless potential and
$\gamma = k_B T/D$ is the friction coefficient, with $k_B$ being the
Boltzmann's constant and $T$ the absolute temperature.  We emphasize
that the drift is time-independent while thermal fluctuations are
isotropic and independent of both time and coordinates.  
%In Sec. \ref{sec:discussion}, we will explain the choice of this setting.

As said earlier, most physical works on restricted diffusion skip the
stochastic differential equation (\ref{eq:Skorokhod}) and deal
directly with the probability density of the position $\X_t$ (also
known as the propagator), $G_q(\x,t|\x_0) d\x = \P_{\x_0}\{ \X_t \in
(\x,\x+d\x)\}$, which obeys the Fokker-Planck equation \cite{Risken}
\begin{equation}  \label{eq:FP}
\partial_t G_q(\x,t|\x_0) = \L_{\x} G_q(\x,t|\x_0) , 
\end{equation}
where
\begin{equation}
\L_{\x} = - (\nabla_{\x} \cdot \mmu(\x)) + D \Delta_{\x}
\end{equation}
is the Fokker-Planck operator.
%(the meaning of the subscript $q$ will be explained below).  
Setting the probability flux density
\begin{equation}
\J_q(\x,t|\x_0) = \mmu(\x) G_q(\x,t|\x_0) - D \nabla_{\x} G_q(\x,t|\x_0),
\end{equation}
the Fokker-Planck equation can be understood as the continuity
equation expressing the probability conservation: $\partial_t
G_q(\x,t|\x_0) = - (\nabla \cdot \J_q)$.

The equation (\ref{eq:FP}) has to be completed by the initial
condition $G_q(\x,0|\x_0) = \delta(\x-\x_0)$ stating that the particle
has started from $\x_0$ at time $t=0$, and by an appropriate boundary
condition that accounts for interactions with the boundary.  When the
boundary is inert and impermeable for the particle, the probability
flux density in the normal direction to the boundary,
\begin{eqnarray} \nonumber
j_q(\s,t|\x_0) &=& \bigl(\n(\x) \cdot \J_q(\x,t|\x_0)\bigr)|_{\x=\s} \\     \label{eq:jinf}
&=& (\n(\s) \cdot \mmu(\s)) G_q(\s,t|\x_0) - D (\partial_n G_q(\x,t|\x_0))|_{\x=\s} ,
\end{eqnarray}
is zero at any boundary point $\s\in\pa$, with $\partial_n = (\n(\x)
\cdot \nabla)$ being the normal derivative oriented outwards the
domain $\Omega$.  In turn, a partially reactive boundary is often
described by {\it imposing} that the probability flux density in the
normal direction is proportional to $G_q(\x,t|\x_0)$,
\begin{equation}  \label{eq:BC_forward}
j_q(\s,t|\x_0) = \kappa \, G_q(\s,t|\x_0) \qquad (\s\in \pa),
\end{equation}
with $\kappa$ (in units m/s) being the reactivity.  This Robin-type
boundary condition was introduced by Collins and Kimball in the
context of chemical physics and was later investigated and employed by
many authors
\cite{Collins49,Sano79,Shoup82,Zwanzig90,Sapoval94,Filoche99,Benichou00,Grebenkov03,Berezhkovskii04,Grebenkov06,Grebenkov06a,Reingruber09,Lawley15,Grebenkov17,Bernoff18b,Grebenkov19}.
When $\kappa =0$, one retrieves the fully reflecting boundary, whereas
in the opposite limit $\kappa \to \infty$, the left-hand side of this
relation vanishes, and one gets the Dirichlet boundary condition
$G(\x,t|\x_0)|_{\x\in \pa} = 0$ for a perfectly reactive boundary.  As
discussed below, the Robin boundary condition (\ref{eq:BC_forward})
describes one surface reaction mechanism among many others.

It is well known that the propagator also satisfies the backward
Fokker-Planck equation \cite{Risken}, which reads in our case as
\begin{equation}  \label{eq:FP_backward}
\partial_t G_q(\x,t|\x_0) = \L_{\x_0}^\dagger G_q(\x,t|\x_0) , 
\end{equation}
where
\begin{equation}
\L_{\x_0}^\dagger = (\mmu(\x_0) \cdot  \nabla_{\x_0}) + D \Delta_{\x_0}
\end{equation}
is the adjoint Fokker-Planck operator.  Since both $\mmu(\x)$ and $D$
are time-independent, solutions of both Fokker-Planck equations depend
on the difference between the terminal and starting times $t$ and $t_0
= 0$, that allowed us to replace the time derivative $-\partial_{t_0}$
by $\partial_t$ in Eq. (\ref{eq:FP_backward}).

The very definition of the propagator $G_q(\x,t|\x_0)$ indicates the
deliberate choice of focusing on the position $\X_t$ and ignoring the
boundary local time $\ell_t$ in the second description.  As shown in
\cite{Grebenkov20} and discussed below, the inclusion of boundary
encounters into the theoretical framework brings numerous advantages.

\subsection{Full propagator}

We start with restricted diffusion without any surface reaction and
introduce the full propagator $P(\x,\ell,t|\x_0)$ as the joint
probability density of $\X_t$ and $\ell_t$ in a bounded domain
$\Omega$ with an {\it impermeable} boundary $\pa$:
\begin{equation}
P(\x,\ell,t|\x_0) d\x d\ell = \P_{\x_0}\{ \X_t\in(\x,\x+d\x),~ \ell_t \in (\ell,\ell+d\ell)\}.  
\end{equation}
In this way, the full propagator is constructed to describe
exclusively the diffusive dynamics in the bulk.  As the
stochastic solution of the Skorokhod equation (\ref{eq:Skorokhod}) is
unique \cite{Ito,Freidlin}, the full propagator is well-defined.

In order to incorporate surface reactions, we exploit the microscopic
interpretation of the boundary condition (\ref{eq:BC_forward}) by
introducing a thin reactive layer of width $a$ near the boundary and
counting the number of times, $\N_t^a$, that the diffusing particle
has crossed this layer up to time $t$.  Each crossing can be
understood as an encounter with the boundary during which the particle
might interact with it.  The self-similar character of Brownian motion
implies that after the first encounter with the boundary, the particle
returns infinitely many times to that boundary \cite{Morters}.  As a
consequence, the number $\N_t^a$ diverges as $a\to 0$, but its
rescaling by $a$ yields a nontrivial limit -- the boundary local time:
\begin{equation}
\ell_t = \lim\limits_{a\to 0} a\N_t^a . 
\end{equation}
This representation is equivalent to Eq. (\ref{eq:ellt_def}) and
yields an alternative approximation to the boundary local time.

At each encounter, the particle can react with the boundary with some
probability $p_a \simeq a \kappa/D$ or resume its diffusion in the
bulk \cite{Filoche99,Grebenkov03,Grebenkov07a}.  Such reaction
attempts are supposed to be independent from each other and from the
diffusion process.  Denoting by $\T$ the time of the successful
reaction event, one gets in the limit $a\to 0$: 
\begin{equation} \fl \label{eq:stopping}
\P\{ \T > t\} = \E\{(1-p_a)^{\N_t^a}\} \approx \E\{(1-a\kappa/D)^{\ell_t/a}\} \approx \E\{e^{-(\kappa/D)\ell_t}\} = \P\{ \ell_t < \hat{\ell}\} \,,
%\P\{ \T > t\} = (1-p_a)^{\N_t^a} \approx (1-a\kappa/D)^{\ell_t/a} \approx e^{-(\kappa/D)\ell_t} = \P\{ \ell_t < \hat{\ell}\} \,,
\end{equation}
where we introduced an independent random threshold $\hat{\ell}$
obeying the exponential law:
\begin{equation}  \label{eq:hat_ell}
\P\{\hat{\ell} > \ell\} = e^{-q\ell} , \qquad \textrm{with}~ q = \kappa/D.
\end{equation}
In other words, the successful reaction event occurs when the boundary
local time $\ell_t$ crosses the random threshold $\hat{\ell}$.  As
first suggested in \cite{Grebenkov06}, this relation can actually be
used as the definition of the first-reaction time $\T$:
\begin{equation}  \label{eq:T_def}
\T = \inf\{ t > 0 ~:~ \ell_t > \hat{\ell}\} ,
\end{equation}
in analogy with the definition of the first-passage time: $\T_{\rm
FPT} = \inf\{ t > 0 ~:~ \X_t \in \pa\}$.  Moreover, as the boundary
local time remains zero until the first encounter, the first-passage
time can also be defined as $\T_{\rm FPT} = \inf\{ t > 0 ~:~ \ell_t >
0\}$, i.e., as the first crossing of the zero threshold.  The boundary
local time offers thus a unified way of treating perfectly and
partially reactive boundaries.

By definition, the conventional propagator $G_q(\x,t|\x_0)$ is the
probability density of finding the particle in a vicinity of $\x$ at
time $t$, given that the particle has started at $\x_0$ and not
reacted on the boundary up to time $t$.  According to
Eq. (\ref{eq:stopping}), the survival of the particle means that
$\ell_t < \hat{\ell}$.  As a consequence, the conventional propagator is
given by the full propagator $P(\x,\ell,t|\x_0)$, multiplied by the
survival probability of that particle up to time $t$, $\P\{\T > t\} =
\P\{\ell_t < \hat{\ell}\}$, and integrated over all possible values of
the boundary local time:
\begin{equation}  \label{eq:Gq_P}
G_q(\x,t|\x_0) = \int\limits_0^\infty d\ell \,\underbrace{e^{-q\ell}}_{=\P\{\ell < \hat{\ell}\}} \, P(\x,\ell,t|\x_0).
\end{equation}
The parameter $q$ stands in the subscript of the propagator to
highlight its dependence on $q$ through the boundary condition
(\ref{eq:BC_forward}).  In other words, the propagator for partially
reactive boundary is obtained as the Laplace transform (with respect
to $\ell$) of the full propagator for purely reflecting boundary.
This relation was established in \cite{Grebenkov20} for ordinary
diffusion without drift.  Whenever surface reactions are independent
from bulk dynamics, the above arguments can be applied, in particular,
Eq. (\ref{eq:Gq_P}) holds for more general diffusion processes with a
drift.  Note also that this relation can also be deduced from the
rigorous probabilistic analysis of a more general Robin boundary value
problem presented in \cite{Papanicolaou90}.  We emphasize that the
boundary local time $\ell_t$ depends exclusively on the diffusive
dynamics in the confining domain $\Omega$ with a purely reflecting
boundary, whereas the threshold $\hat{\ell}$ depends exclusively on
the reactivity $q$, allowing one to disentangle these two aspects of
diffusion-controlled reactions.  This disentanglement opens a way to
investigate much more elaborate surface reaction mechanisms when the
exponential distribution of the random threshold $\hat{\ell}$ is
replaced by another distribution (see \cite{Grebenkov20} for details).
As a consequence, the unique full propagator $P(\x,\ell,t|\x_0)$ can
determine a variety of conventional propagators by choosing an
appropriate stopping condition.  In summary, the stochastic foundation
(\ref{eq:Skorokhod}) for both conventional and encounter-based
approaches is the same, but the ways of incorporating surface
reactions are different (Robin boundary condition versus random
threshold $\hat{\ell}$).

While the full propagator $P(\x,\ell,t|\x_0)$ determines any
conventional propagator $G_q(\x,t|\x_0)$ via Eq. (\ref{eq:Gq_P}), its
inverse Laplace transform with respect to $q$ can in turn be used to
access the full propagator.  However, the implicit dependence of the
conventional propagator on $q$ as the parameter in the boundary
condition (\ref{eq:BC_forward}) makes difficult further explorations
of such an inverse, even numerically.  To overcome this limitation, a
spectral representation of the full propagator was derived in
\cite{Grebenkov20} for ordinary diffusion.  Here we aim at extending
this representation to restricted diffusion with a gradient drift.

\subsection{Spectral decompositions}

In the following, we mainly deal with Laplace-transformed quantities
with respect to time $t$, denoted by tilde.  For instance,
\begin{equation}
\tilde{G}_q(\x,p|\x_0) = \int\limits_0^\infty dt \, e^{-pt} \, G_q(\x,t|\x_0)
\end{equation}
is the Green's function for the Fokker-Planck equation, which
admits both forward and backward forms:
\begin{equation}  \label{eq:Gq_fFP}
\left\{ \begin{array}{l l} (p - \L_{\x}) \tilde{G}_q(\x,p|\x_0) = \delta(\x-\x_0) & (\x\in\Omega)  \\  \label{eq:Gq_fFP_BC}
\tilde{j}_q(\x,p|\x_0) = \kappa \tilde{G}_q(\x,p|\x_0) & (\x\in\pa) \\ \end{array} \right. \quad \textrm{(forward)}
\end{equation}
and
\begin{equation}
\left\{ \begin{array}{ll} (p - \L^\dagger_{\x_0}) \tilde{G}_q(\x,p|\x_0) = \delta(\x-\x_0) & (\x_0\in\Omega)  \\
\partial_{n_0} \tilde{G}(\x,p|\x_0) + q \tilde{G}_q(\x,p|\x_0) = 0 & (\x_0\in\pa) \\ \end{array} \right.  \quad \textrm{(backward)}
\end{equation}
where  
\begin{equation}  \label{eq:j}
\tilde{j}_q(\x,p|\x_0) 
= (\n(\x) \cdot \mmu(\x)) \tilde{G}_q(\x,p|\x_0) - D \partial_n \tilde{G}_q(\x,p|\x_0) \quad (\x\in\pa)
\end{equation}
is the Laplace-transformed probability flux density.

In \cite{Grebenkov20}, we considered ordinary restricted diffusion
(without drift) and employed the so-called Dirichlet-to-Neumann
operator $\M_p$ that associates to a given function $f$ on the
boundary $\pa$ another function on that boundary such that $\M_p f =
(\partial_n u)_{|\pa}$, where $u(\x)$ satisfies $(p-D\Delta)u(\x) = 0$
in $\Omega$ and $u_{|\pa} = f$.  In other words, $\M_p$ maps the
Dirichlet boundary condition onto Neumann boundary condition.  Relying
on the fact that the self-adjoint operator $\M_p$ on a bounded
boundary has a discrete spectrum (see
\cite{Arendt14,Daners14,Elst14,Berhndt15,Arendt15,Hassell17,Girouard17} for
mathematical details), we derived the following spectral decomposition
\begin{equation}  \label{eq:Gq_spectral0}
\tilde{G}_q(\x,p|\x_0) = \tilde{G}_\infty(\x,p|\x_0) + \frac{1}{D} \sum\limits_n \frac{V_n^{(p)}(\x) [V_n^{(p)}(\x_0)]^*}{q + \mu_n^{(p)}} \,,
\end{equation}
where $\mu_n^{(p)}$ are the eigenvalues of $\M_p$, while
$V_n^{(p)}(\x)$ are projections of $\tilde{j}_\infty(\x,p|\x_0)$ onto
the eigenfunctions $v_n^{(p)}$ of that operator (see below for
details; note that Eq. (\ref{eq:Gq_spectral0}) is not used in the
following derivation and is reproduced here just for motivation).  The
inverse Laplace transform of this relation with respect to $q$ yielded
the spectral decomposition of the Laplace-transformed full propagator:
\begin{equation}  \label{eq:P_spectral0}
\tilde{P}_q(\x,\ell,p|\x_0) = \tilde{G}_\infty(\x,p|\x_0)\delta(\ell) + 
\frac{1}{D} \sum\limits_n V_n^{(p)}(\x) [V_n^{(p)}(\x_0)]^* \, e^{-\ell \mu_n^{(p)}} \,.
\end{equation}

A naive extension of this approach to restricted diffusion with drift
would fail.  In fact, even though an appropriate Dirichlet-to-Neumann
operator could be introduced by replacing the equation
$(p-D\Delta)u=0$ by $(p-\L_{\x})u = 0$ or $(p-\L^\dagger_{\x_0})u =
0$, such an extension would not in general be self-adjoint and thus
would require much more elaborate spectral tools.

To overcome this limitation, we reformulate the problem in terms of
the symmetrized Fokker-Planck operator \cite{Risken}
\begin{equation}  \label{eq:Ls}
\bar{\L} = D e^{\frac12 \Phi(\x)} \nabla_{\x} e^{-\Phi(\x)} \nabla_{\x} e^{\frac12 \Phi(\x)} 
= D \Delta - \left(\frac{(\nabla \cdot \mmu(\x))}{2} + \frac{|\mmu(\x)|^2}{4D}\right) \,,
\end{equation}
where the second term in the last relation is the multiplication by
the explicit drift-dependent function (here, the operator $\nabla$
acts only on $\mmu(\x)$).  To avoid mathematical subtleties, we assume
that the second term is a smooth bounded function on $\Omega$.  With
the aid of this expression, the Fokker-Planck operator and its adjoint
can be written as
\begin{eqnarray}
\L &=& e^{-\frac12\Phi(\x)} \bar{\L} e^{\frac12\Phi(\x)} = D \nabla_{\x} e^{-\Phi(\x)} \nabla_{\x} e^{\Phi(\x)},  \\
\L^\dagger &=& e^{\frac12\Phi(\x)} \bar{\L} e^{-\frac12 \Phi(\x)} = D e^{\Phi(\x)} \nabla_{\x} e^{-\Phi(\x)}  \nabla_{\x} .
\end{eqnarray}
In order to reduce the original problem to that with $\bar{\L}$, one
can represent the Green's function as
\begin{equation}  \label{eq:Gq_gq}
\tilde{G}_q(\x,p|\x_0) = e^{\frac12 \Phi(\x_0)- \frac12 \Phi(\x)} \tilde{g}_q(\x,p|\x_0),
\end{equation}
so that forward and backward equations imply two equivalent boundary
value problems for the new function $\tilde{g}_q(\x,p|\x_0)$:
\begin{eqnarray}    \label{eq:FP_gq}
(p - \bar{\L}_{\x}) \tilde{g}_q(\x,p|\x_0) = \delta(\x-\x_0)  &&\quad (\x\in\Omega), \\
\bigl(\partial_{n} + q + \phi(\x)\bigr) \tilde{g}_q(\x,p|\x_0) = 0 && \quad (\x \in \pa) ,
\end{eqnarray}
and
\begin{eqnarray} 
(p - \bar{\L}_{\x_0}) \tilde{g}_q(\x,p|\x_0) = \delta(\x-\x_0) && \quad (\x_0\in\Omega), \\
\bigl(\partial_{n_0} + q + \phi(\x_0)\bigr) \tilde{g}_q(\x,p|\x_0) = 0 && \quad (\x_0\in \pa),
\end{eqnarray}
with
\begin{equation}
\phi(\x) = \frac12 (\partial_{n} \Phi(\x)) = - \frac{1}{2D} \bigl(\n(\x) \cdot \mmu(\x)\bigr) \qquad (\x\in \pa).
\end{equation}
As the forward and backward problems are now identical, their solution
is symmetric with respect to the exchange of $\x$ and $\x_0$:
$\tilde{g}_q(\x,p|\x_0) = \tilde{g}_q(\x_0,p|\x)$.

We search for the following representation:
\begin{equation}  \label{eq:Gq_U_auxil2}
\tilde{g}_q(\x,p|\x_0) = \tilde{g}_\infty(\x,p|\x_0) + \frac{1}{D} u(\x,p|\x_0),
\end{equation}
where the new function $u(\x,p|\x_0)$ satisfies
\begin{equation}  
(p - \bar{\L}_{\x_0}) u(\x,p|\x_0) = 0 ,
\end{equation}
subject to the boundary condition at $\x_0\in \pa$:
\begin{eqnarray*}  \fl
0 &=& (\partial_{n_0} + q)\tilde{G}_q(\x,p|\x_0) = (\partial_{n_0} + q)  \biggl(\tilde{G}_\infty(\x,p|\x_0) 
+ \frac{e^{\frac12 \Phi(\x_0)- \frac12\Phi(\x)}}{D} u(\x,p|\x_0) \biggr) \\ \fl
&=& \partial_{n_0} \tilde{G}_\infty(\x,p|\x_0) + \frac{e^{\frac12 \Phi(\x_0)- \frac12\Phi(\x)}}{D} 
\bigl(\partial_{n_0} + q + \phi(\x_0)\bigr) u(\x,p|\x_0),
\end{eqnarray*}
i.e.,
\begin{equation}  \label{eq:BC_auxil}
\bigl(\partial_{n_0} + q + \phi(\x_0)\bigr) u(\x,p|\x_0) = e^{\frac12 \Phi(\x)-\frac12 \Phi(\x_0)}  \tilde{j}'_\infty(\x_0,p|\x) 
\quad (\x_0 \in \pa),
\end{equation}
where 
\begin{equation}  \label{eq:jprime}
\tilde{j}'_{\infty}(\x_0,p|\x) = -D \partial_{n_0} \tilde{G}_\infty(\x,p|\x_0) \qquad (\x_0 \in \pa).
\end{equation}
Since the normal derivative $\partial_{n_0}$ acts on $\x_0$ of
$\tilde{G}_q(\x,p|\x_0)$, this function differs in general from
$\tilde{j}_{\infty}(\x,p|\x_0)$ given in Eq. (\ref{eq:j}).  We also
emphasize that the order of $\x_0$ and $\x$ in this notation has been
changed, in order to keep the convention that the first argument of
$j$ is a point on the boundary.

In the next step, we introduce a Dirichlet-to-Neumann operator $\M_p$
that associates to a given function $f$ on $\pa$ another function on
$\pa$ such that
\begin{equation}  \label{eq:DtN}
\M_p f = \bigl(\partial_n w + \phi(\x) w\bigr)_{|\pa},  \quad \textrm{with} ~~ 
\left\{ \begin{array}{ll} (p - \bar{\L}_{\x}) w = 0 & (\x\in \Omega), \\  w = f & (\x\in \pa). \\ \end{array} \right.
\end{equation}
The operator $\bar{\L}$ in Eq. (\ref{eq:Ls}) having the form of a
Schr\"odinger operator, is self-adjoint; in addition, as the domain is
bounded, the drift-dependent term is a bounded perturbation to the
Laplace operator.  In analogy with the Laplacian case (see
\cite{Elst14,Berhndt15} for mathematical details), one can expect
that, under mild assumptions on the drift term, the
Dirichlet-to-Neumann operator $\M_p$ is self-adjoint, with a discrete
spectrum of real eigenvalues $\{\mu_n^{(p)}\}$, while its
eigenfunctions $\{ v_n^{(p)}\}$ form a complete orthonormal basis in
$L_2(\pa)$.  As a proof of this mathematical statement is beyond the
scope of the present paper, we will use it as a conjecture in the
following presentation.  In other words, we focus on such drifts
$\mmu(\x)$, for which this statement is valid.
The Dirichlet-to-Neumann operator $\M_p$ allows us to rewrite the
boundary condition (\ref{eq:BC_auxil}) in an operator form, from which
\begin{equation}  %\label{eq:BC_auxil}
u(\x,p|\x_0) = (\M_p + q)^{-1} e^{\frac12 \Phi(\x)- \frac12 \Phi(\x_0)}  \tilde{j}'_\infty(\x_0,p|\x)    \quad (\x_0 \in \pa).
\end{equation}

Finally, one needs to extend $u(\x,p|\x_0)$ to any $\x_0 \in \Omega$.
For this purpose, one multiplies Eq. (\ref{eq:FP_gq}) with $q =
\infty$ by $u(\x,p|\x')$ and subtracts from it the equation
$(p-\bar{\L}_{\x})u(\x,p|\x') = 0$ multiplied by
$\tilde{g}_\infty(\x',p|\x_0)$.  The integration of both parts over
$\x'$ yields
\begin{eqnarray*} \fl
u(\x,p|\x_0) &=& \int\limits_{\Omega} d\x' \, u(\x,p|\x') \delta(\x'-\x_0) \\ \fl 
&=& 
\int\limits_{\Omega} d\x' \biggl(u(\x,p|\x') (p - \bar{\L}_{\x'}) \tilde{g}_\infty(\x',p|\x_0) - 
\tilde{g}_\infty(\x',p|\x_0) (p - \bar{\L}_{\x'}) u(\x,p|\x')\biggr) \\ \fl
&=& -
\int\limits_{\Omega} d\x' \biggl(u(\x,p|\x') D \Delta_{\x'} \tilde{g}_\infty(\x',p|\x_0) - 
\tilde{g}_\infty(\x',p|\x_0) D \Delta_{\x'}) u(\x,p|\x')\biggr) \\ \fl
&=& - \int\limits_{\pa} d\x' \biggl(u(\x,p|\x') D \partial_{n'} \tilde{g}_\infty(\x',p|\x_0) -
\tilde{g}_\infty(\x',p|\x_0) D \partial_{n'} u(\x,p|\x') \biggr) \\ \fl
&=& \int\limits_{\pa} d\x' u(\x,p|\x') e^{\frac12 \Phi(\x')-\frac12 \Phi(\x_0)} \partial_{n'} (-D \tilde{G}_\infty(\x',p|\x_0)) \\ \fl
&=& \int\limits_{\pa} d\x' u(\x,p|\x') e^{\frac12 \Phi(\x')-\frac12 \Phi(\x_0)} \tilde{j}_\infty(\x',p|\x_0) ,
\end{eqnarray*}
where we used Green's relations and the Dirichlet boundary condition
$\tilde{G}_\infty(\x,p|\x_0) = 0$ at $\x\in\pa$.  As a consequence, we
get
\begin{eqnarray} \fl   \nonumber
&& \tilde{G}_q(\x,p|\x_0) = \tilde{G}_\infty(\x,p|\x_0) + \frac{e^{\frac12 \Phi(\x_0)- \frac12 \Phi(\x)}}{D} u(\x,p|\x_0) \\ \fl  \nonumber
&& \quad = \tilde{G}_\infty(\x,p|\x_0) + \frac{e^{- \frac12 \Phi(\x)}}{D} 
\int\limits_{\pa} d\x' u(\x,p|\x') e^{\frac12 \Phi(\x')} \tilde{j}_\infty(\x',p|\x_0) \\ \fl  \label{eq:Gq_auxil3}
&& \quad = \tilde{G}_\infty(\x,p|\x_0) + \frac{1}{D} 
\int\limits_{\pa} d\x' \biggl(\bigl(\M_p + q\bigr)^{-1} e^{- \frac12 \Phi(\x')}  \tilde{j}'_\infty(\x',p|\x)\biggr) 
e^{\frac12 \Phi(\x')} \tilde{j}_\infty(\x',p|\x_0).
\end{eqnarray}
Using the eigenfunctions of $\M_p$, one gets the spectral
decomposition of the Green's function:
\begin{equation}  \label{eq:Gq_spectral}
\tilde{G}_q(\x,p|\x_0) = \tilde{G}_\infty(\x,p|\x_0) + \frac{1}{D} \sum\limits_n \frac{V^{'(p)}_n(\x) [V^{(p)}_n(\x_0)]^*}{q + \mu_n^{(p)}} \,,
\end{equation}
where
\begin{eqnarray}  \label{eq:Vn}
V^{(p)}_n(\x_0) &=& \int\limits_{\pa} d\s \, v_n^{(p)}(\s) \, e^{\frac12 \Phi(\s)} \tilde{j}_\infty(\s,p|\x_0) ,\\  \label{eq:Vpn}
V^{'(p)}_n(\x) &=& \int\limits_{\pa} d\s \, v_n^{(p)}(\s) \, e^{-\frac12 \Phi(\s)} \tilde{j}'_\infty(\s,p|\x) .
\end{eqnarray}
When there is no drift, one has $\Phi(\s) = 0$ and thus
$V^{'(p)}_n(\x) = V^{(p)}_n(\x)$, so that Eq. (\ref{eq:Gq_spectral})
is reduced to the former result (\ref{eq:Gq_spectral0}) derived in
\cite{Grebenkov20}.  The Laplace inversion of
Eq. (\ref{eq:Gq_spectral}) with respect to $q$ yields
\begin{equation}  \label{eq:P_spectral}
\tilde{P}(\x,\ell,p|\x_0) = \tilde{G}_\infty(\x,p|\x_0) \delta(\ell) 
+ \frac{1}{D} \sum\limits_n V^{'(p)}_n(\x) [V^{(p)}_n(\x_0)]^* e^{-\ell \mu_n^{(p)}} \,.
\end{equation}
This is our main theoretical result, which generalizes the former
relation (\ref{eq:P_spectral0}) to the case of restricted diffusion
with a gradient drift.

\subsection{Various diffusion-reaction characteristics}

As said earlier, the full propagator determines various
characteristics of diffusion-influenced reactions.  For instance,
integrating Eqs. (\ref{eq:Gq_spectral}, \ref{eq:P_spectral}) over $\x
\in\Omega$, one accesses the Laplace-transformed survival probability and
the Laplace-transformed probability density of the boundary local
time, respectively:
\begin{equation}  \label{eq:Sq_spectral}
\tilde{S}_q(p|\x_0) = \tilde{S}_\infty(p|\x_0) + \frac{1}{D} \sum\limits_n \frac{\overline{V^{'(p)}_n} [V^{(p)}_n(\x_0)]^*}{q + \mu_n^{(p)}} \,,
\end{equation}
and
\begin{equation}  \label{eq:Pell_spectral}
\tilde{P}(\circ,\ell,p|\x_0) = \tilde{S}_\infty(p|\x_0) \delta(\ell) 
+ \frac{1}{D} \sum\limits_n \overline{V^{'(p)}_n} [V^{(p)}_n(\x_0)]^* e^{-\ell \mu_n^{(p)}} \,,
\end{equation}
where $\circ$ denotes the marginalized variable $\x$, and
\begin{equation}
\overline{V^{'(p)}_n} = \int\limits_{\Omega} d\x \, V^{'(p)}_n(\x).
\end{equation}
In \ref{sec:auxil}, we derive the following representation for this
quantity:
\begin{equation}    \label{eq:Vn_int} 
\overline{V_n^{'(p)}} = \frac{D}{p} \mu_n^{(p)} \int\limits_{\pa} d\s \, e^{-\frac12\Phi(\s)} v_n^{(p)}(\s).
\end{equation}
These expressions generalize our former results for ordinary diffusion
without drift from \cite{Grebenkov19b,Grebenkov20c}.

We recall that the survival probability determines the distribution of
the first-reaction time $\T$ on a partially reactive boundary
\cite{Redner}: $S_q(t|\x_0) = \P_{\x_0}\{ \T > t\}$.  At the
same time, $S_q(t|\x_0) = \E_{\x_0}\{ e^{-q\ell_t} \}$ is the
generating function of the boundary local time $\ell_t$
\cite{Grebenkov20,Grebenkov19} that allows one to compute its moments
as
\begin{equation}
\E_{\x_0}\{ \ell_t^k\} = (-1)^k \lim\limits_{q\to 0} \frac{\partial^k}{\partial q^k} S_q(t|\x_0),
\end{equation}
from which
\begin{equation}  \label{eq:ellt_moments}
\int\limits_0^\infty dt\, e^{-pt} \, \E_{\x_0}\{ \ell_t^k\} =  \frac{k!}{D} 
\sum\limits_n \frac{\overline{V^{'(p)}_n} [V^{(p)}_n(\x_0)]^*}{[\mu_n^{(p)}]^{k+1}} .
\end{equation}
Multiplying both sides by $p$, one can also interpret the integral
over $t$ as the double average of the random variable $\ell_\tau^k$
with the random exponentially distributed stopping time $\tau$.
Similarly, $p \tilde{P}(\circ, \ell,p|\x_0)$ can be interpreted as the
probability density of $\ell_\tau$:
\begin{equation} \fl
\P_{\x_0}\{ \ell_\tau \in (\ell,\ell+d\ell)\} = \int\limits_0^\infty dt \, \underbrace{p\, e^{-pt}}_{\textrm{\tiny pdf of}~\tau} 
\, \P_{\x_0}\{ \ell_t \in (\ell,\ell+d\ell)\} = p \tilde{P}(\circ, \ell,p|\x_0) d\ell .
\end{equation}

Setting $q = 0$ in Eq. (\ref{eq:Sq_spectral}) and noting that
$\tilde{S}_0(p|\x_0) = 1/p$ (since $S_0(t|\x_0) = 1$ for a fully inert
boundary), one gets
\begin{equation}  \label{eq:Sinf_spectral}
\tilde{S}_\infty(p|\x_0) = \frac{1}{p} - \frac{1}{D} \sum\limits_n \frac{\overline{V^{'(p)}_n} [V^{(p)}_n(\x_0)]^*}{\mu_n^{(p)}} \,.
\end{equation}
Substituting this expression back into Eq. (\ref{eq:Sq_spectral}), one
deduces an equivalent spectral decomposition:
\begin{equation}  \label{eq:Sq_spectral2}
\tilde{S}_q(p|\x_0) = \frac{1}{p} - \frac{1}{D} \sum\limits_n \frac{\overline{V^{'(p)}_n} [V^{(p)}_n(\x_0)]^*}{\mu_n^{(p)}(1 + \mu_n^{(p)}/q)} \,,
\end{equation}
from which one also derives the Laplace-transformed probability
density of the first-reaction time:
\begin{equation}  \label{eq:Hq_spectral}
\tilde{H}_q(p|\x_0) = 1 - p \tilde{S}_q(p|\x_0) 
= \frac{p}{D} \sum\limits_n \frac{\overline{V^{'(p)}_n} [V^{(p)}_n(\x_0)]^*}{\mu_n^{(p)}(1 + \mu_n^{(p)}/q)} \,.
\end{equation}

As $H_q(t|\x_0)$ can be interpreted as the probability flux of
particles, started from $\x_0$, onto the reactive boundary, the total
flux $J_q(t)$ is obtained by averaging $H_q(t|\x_0)$ with the initial
concentration $c(\x_0)$ of these particles.  If the initial
concentration is uniform, $c(\x_0) = c_0$, one gets in the Laplace
domain:
\begin{equation}  \label{eq:Jq_spectral}
\tilde{J}_q(p) = \int\limits_{\Omega} d\x_0 \, c_0 \, \tilde{H}_q(p|\x_0)
= c_0 \frac{p}{D} \sum\limits_n \frac{\overline{V^{'(p)}_n} [\overline{V^{(p)}_n}]^*}{\mu_n^{(p)}(1 + \mu_n^{(p)}/q)} \,,
\end{equation}
where
\begin{equation}
\overline{V^{(p)}_n} = \int\limits_{\Omega} d\x_0 \, V^{(p)}_n(\x_0) .
\end{equation}

Finally, the definition (\ref{eq:T_def}) of the first-reaction
time $\T$ puts forward the importance of threshold crossing by the
boundary local time $\ell_t$.  Following \cite{Grebenkov20}, let us
consider a fixed threshold $\ell$ and investigate its first-crossing
time $\T_\ell = \inf\{ t>0~:~ \ell_t > \ell\}$ (note that $\T =
\T_{\hat{\ell}}$, i.e., when a fixed threshold $\ell$ is replaced by a
random threshold $\hat{\ell}$).  As the boundary local time is a
non-decreasing process, one has $\P\{\T_\ell > t\} = \P\{\ell_t <
\ell\}$, from which the probability density $U(\ell,t|\x_0)$ of
$\T_\ell$ can be expressed as
\begin{equation}
U(\ell,t|\x_0) = - \partial_t \P\{\T_\ell > t\} = \partial_t \P\{\ell_t > \ell\} 
= \partial_t \int\limits_{\ell}^\infty d\ell' \, P(\circ,\ell',t|\x_0) .
\end{equation}
In the Laplace domain, the substitution of Eq. (\ref{eq:Pell_spectral})
yields
\begin{equation}
\tilde{U}(\ell,p|\x_0) = \frac{p}{D} \sum\limits_n \frac{\overline{V^{'(p)}_n}}{\mu_n^{(p)}} [V^{(p)}_n(\x_0)]^* e^{-\ell \mu_n^{(p)}} \,.
\end{equation}
As $\T = \T_{\hat{\ell}}$, the probability densities of $\T$ and
$\T_\ell$ are related:
\begin{eqnarray*} 
H_q(t|\x_0)  &=& \frac{d\P\{\T\in(t,t+dt)\}}{dt} 
= \int\limits_0^\infty d\ell \, q e^{-q\ell} \, \frac{d\P\{\T_{\ell}\in(t,t+dt)\}}{dt} \\
&=& \int\limits_0^\infty d\ell \, q e^{-q\ell} \, U(\ell,t|\x_0)  ,
\end{eqnarray*}
where $qe^{-q\ell}$ is the probability density of the exponential
threshold $\hat{\ell}$.  As the full propagator determines a variety
of conventional propagators, the probability density $U(\ell,t|\x_0)$
determines a variety of the first-reaction time densities.

\subsection{Probabilistic interpretation}

To conclude this section, we provide some probabilistic insights onto
the above results.  For this purpose, we write an alternative
representation of the Green's function:
\begin{equation} \fl  \label{eq:auxil56}
\tilde{G}_q(\x,p|\x_0) = \tilde{G}_\infty(\x,p|\x_0) + \int\limits_{\pa} d\s_1 
\int\limits_{\pa} d\s_2 \, \tilde{j}_\infty(\s_1,p|\x_0) \tilde{G}_q(\s_2,p|\s_1) \tilde{j}'_\infty(\s_2,p|\x).
\end{equation}
This relation claims that either a random trajectory starting from
$\x_0$ goes directly to $\x$ without hitting the boundary (the first
term), or this trajectory hits the boundary at least once before
arriving at $\x$ (the second term).  In the latter case, the Markov
property of restricted diffusion implies the product of three
contributions in the Laplace domain: the particle hits the boundary
for the first time at some point $\s_1$ (the factor
$\tilde{j}_\infty(\s_1,p|\x_0)$), moves inside the domain until the
last hit of the boundary at some point $\s_2$ (the factor
$\tilde{G}_q(\s_2,p|\s_1)$), and finally goes directly to the bulk
point $\x$ (the factor $\tilde{j}'_\infty(\s_2,p|\x)$).  A direct
definition of the last factor would require conditioning Brownian
trajectories to avoid hitting the boundary.  A simpler approach
consists in the simultaneous time and drift reversal (i.e., $\mmu(\x)
\to -\mmu(\x)$) in the diffusive dynamics, for which
$j'_\infty(\s_2,t|\x)$ can be understood as the probability flux
density onto the absorbing boundary at the point $\s_2$ at time $t$
when starting from $\x$.  In this way, one restores the reversal
symmetry of restricted diffusion, in particular, one gets
$G'_q(\x_0,t|\x) = G_q(\x,t|\x_0)$ and thus
\begin{equation}  \label{eq:jprime2}
\tilde{j}'_{\infty}(\x_0,p|\x) = -D \partial_{n_0} \tilde{G}'_\infty(\x_0,p|\x) .
\end{equation}
When there is no drift, one simply has $G_q(\x_0,t|\x) =
G'_q(\x_0,t|\x) = G_q(\x,t|\x_0)$ and also $j'_\infty(\s,t|\x) =
j_\infty(\s,t|\x)$.

Comparison of Eqs. (\ref{eq:Gq_auxil3}, \ref{eq:auxil56}) implies that
\begin{equation*}
(\M_p + q)_{\s_1} D \underbrace{e^{\frac12 \Phi(\s_2) - \frac12 \Phi(\s_1)} \tilde{G}_q(\s_2,p|\s_1)}_{=\tilde{g}_q(\s_2,p|\s_1)} = \delta(\s_1 - \s_2)
\qquad (\s_1,\s_2 \in \pa),
\end{equation*}
i.e., $D \tilde{g}_q(\s_2,p|\s_1)$ can be understood as the kernel
density of the operator $(\M_p + q)^{-1}$.

\section{Constant drift on an interval}
\label{sec:interval}

In order to illustrate the derived spectral decompositions, we
consider a simple yet informative setting of restricted diffusion on
an interval $(0,L)$ with a constant drift $\mu$.  This problem is
mathematically equivalent to restricted diffusion between parallel
plates given that the lateral motion along the plates does not affect
boundary encounters.  For instance, this mathematical setting can
describe the motion of a charged particle in a constant electric field
inside a capacitor.  The geometric simplicity of the problem will
allow us to get fully explicit formulas and thus to clearly illustrate
the effect of the drift onto the statistics of boundary encounters.
Qualitatively, a drift towards the boundary tends to keep the particle
in a vicinity of the boundary and thus to enlarge the number of
encounters, whereas the reversed drift is expected to result in the
opposite effect.  However, these effects have not been studied
quantitatively, to our knowledge.

The constant drift corresponds to the potential $\Phi(x) = - \mu x/D$.
Expectedly, the Fokker-Planck operator $\L = D\partial_x^2 - \mu
\partial_x$ is not self-adjoint, whereas the symmetric operator, which
takes a simple form
\begin{equation}
\bar{\L} = D \partial_x^2 - \mu^2/(4D) = D(\partial_x^2 - \gamma^2) ,  \qquad \gamma = - \frac{\mu}{2D} \,,
\end{equation}
is self-adjoint.

\subsection{Dirichlet-to-Neumann operator}

As the boundary of an interval consists of two endpoints, the
pseudo-differential operator $\M_p$ is reduced to a $2\times 2$
matrix.  In fact, a general solution of the equation $(p-\bar{\L})u
=0$ has the form $u(x) = c_+ e^{\beta x} + c_- e^{-\beta x}$, with
$\beta = \sqrt{p/D + \gamma^2}$ and two coefficients $c_\pm$ to be
fixed by boundary conditions.  Any ``function'' on the boundary can
thus be obtained as a superposition of two linearly independent
vectors: $f_1 = (1,0)^\dagger$ and $f_2 = (0,1)^\dagger$, where the
first and second components stand for the values at the endpoints $x =
0$ and $x = L$, respectively.  According to Eq. (\ref{eq:DtN}), the
operator $\M_p$ acts as
\begin{equation} 
\M_p \left(\begin{array}{c} u(0) \\ u(L) \\ \end{array} \right) =
\left(\begin{array}{c} (\partial_n u)_{x=0} + \phi(0) u(0) \\ (\partial_n u)_{x=L} + \phi(L) u(L)\\ \end{array} \right)  ,
\end{equation}
where $\phi(0) = - \frac12 (\partial_x \Phi)_{x=0} = \mu/(2D) = -
\gamma$ and $\phi(L) = \frac12 (\partial_x \Phi)_{x=L} = \gamma$.  The
boundary ``function'' $f_1 = (1,0)^\dagger$ is obtained with
\begin{equation}
c_+ = \frac{-1}{e^{2\beta L} - 1} \,, \qquad  c_- = \frac{1}{1-e^{-2\beta L}} \,,
\end{equation}
while $f_2 = (0,1)^\dagger$ corresponds to
\begin{equation}
c_+ = \frac{e^{\beta L}}{e^{2\beta L}-1} \,, \qquad c_- = - \frac{e^{\beta L}}{e^{2\beta L}-1} \,.
\end{equation}
We get then
\begin{eqnarray}  
\M_p & =& \left(\begin{array}{cc} \beta \ctanh(\beta L) - \gamma & - \beta/\sinh(\beta L) \\
-\beta/\sinh(\beta L) & \beta \ctanh(\beta L) + \gamma \\
\end{array} \right).
\end{eqnarray}
Expectedly, this is a Hermitian matrix with two real positive
eigenvalues,
\begin{equation}  \label{eq:mupm}
\mu_{\pm}^{(p)} = \beta \ctanh(\beta L) \pm \sqrt{\gamma^2 + \frac{\beta^2}{\sinh^2(\beta L)}} \,,
\end{equation}
and orthonormal eigenvectors:
\begin{equation}
v_{\pm}^{(p)} = \frac{-1}{\sqrt{\bigl(\mu^{(p)}_\pm - (\M_p)_{11}\bigr)^2 + (\M_p)_{12}^2}} 
\left(\begin{array}{c} (\M_p)_{12} \\ \mu^{(p)}_\pm - (\M_p)_{11} \\ \end{array} \right).
\end{equation}
To distinguish two eigenmodes, here we use the subscripts $+$ and $-$
instead of the index $n$ employed in Sec. \ref{sec:general}.

\subsection{Dirichlet Green's function}

To complete the computation, one needs to evaluate the probability
flux densities.  In \ref{sec:Green}, we recall the derivation of the
Green's function on the interval.  For perfectly reactive endpoints,
one has
\begin{equation} \fl
\tilde{G}_{\infty}(x,p|x_0) = \frac{e^{-\gamma(x-x_0)}}{\beta D \sinh(\beta L)}
\times \left\{ \begin{array}{ll}  \sinh(\beta \bar{x}_0) \sinh(\beta x)  &  (0 < x < x_0), \\
\sinh(\beta \bar{x}) \sinh(\beta x_0) & (x_0 < x < L), \\  \end{array} \right.    
\end{equation}
where $\bar{x} = L - x$ and $\bar{x}_0 = L - x_0$.  The
Laplace-transformed probability flux density reads then
\begin{eqnarray}  \fl
\tilde{j}_\infty(0,p|x_0) &=& \left. \bigg(\mu  \tilde{G}_\infty(x,p|x_0) + D\partial_x \tilde{G}_\infty(x,p|x_0)\biggr)\right|_{x=0} 
=  \frac{e^{\gamma x_0} \sinh(\beta \bar{x}_0)}{\sinh(\beta L)} \,, \\  \fl
\tilde{j}_\infty(L,p|x_0) &=& \left. \bigg(\mu  \tilde{G}_\infty(x,p|x_0) - D\partial_x \tilde{G}_\infty(x,p|x_0)\biggr)\right|_{x=L} 
= \frac{e^{\gamma(x_0-L)} \sinh(\beta x_0)}{\sinh(\beta L)} \,.
\end{eqnarray}
Similarly, we have
\begin{eqnarray}  \fl
\tilde{j}'_\infty(0,p|x) &=& \left. \bigg( D\partial_{x_0} \tilde{G}_\infty(x,p|x_0)\biggr)\right|_{x_0=0} 
=  \frac{e^{-\gamma x} \sinh(\beta \bar{x})}{\sinh(\beta L)} \,, \\  \fl
\tilde{j}'_\infty(L,p|x) &=& \left. \bigg( - D\partial_{x_0} \tilde{G}_\infty(x,p|x_0)\biggr)\right|_{x_0=L} 
= \frac{e^{-\gamma(x-L)} \sinh(\beta x)}{\sinh(\beta L)} \,.
\end{eqnarray}
Expectedly, the functions $\tilde{j}_\infty(0,p|x_0)$ and
$\tilde{j}'_\infty(0,p|x_0)$ differ only by the sign of the drift
(here, the sign of $\gamma$).

\subsection{Propagators}

According to Eqs. (\ref{eq:Vn}, \ref{eq:Vpn}), one has to project the
probability flux densities onto the eigenfunctions of $\M_p$ to get
$V_n^{(p)}$ and $V_n^{'(p)}$.  As the boundary consists of two
endpoints, integrals are reduced to sums with two contributions:
\begin{eqnarray*}
V_\pm^{(p)}(x_0) &=& v_{\pm}^{(p)}(0) \tilde{j}_\infty(0,p|x_0) + v_{\pm}^{(p)}(L) e^{\gamma L} \tilde{j}_\infty(L,p|x_0) \\
&=& \frac{e^{\gamma x_0}}{\sinh(\beta L)}  \biggl(v_{\pm}^{(p)}(0)  \sinh(\beta \bar{x}_0) + v_{\pm}^{(p)}(L) \sinh(\beta x_0)\biggr) , \\
V^{'(p)}_\pm(x) &=& v_{\pm}^{(p)}(0) \tilde{j}'_\infty(0,p|x) + v_{\pm}^{(p)}(L) e^{-\gamma L} \tilde{j}'_\infty(L,p|x) \\
&=& \frac{e^{-\gamma x}}{\sinh(\beta L)} \biggl(v_{\pm}^{(p)}(0) \sinh(\beta \bar{x}) + v_{\pm}^{(p)}(L) \sinh(\beta x) \biggr).
\end{eqnarray*}
As a consequence, Eqs. (\ref{eq:Gq_spectral}, \ref{eq:P_spectral})
imply
\begin{equation} \fl
\tilde{G}_q(x,p|x_0) = \tilde{G}_\infty(x,p|x_0) + \frac{1}{D} \left(\frac{V_+^{(p)}(x_0) V^{'(p)}_+(x)}{q + \mu_+^{(p)}} 
+ \frac{V_-^{(p)}(x_0) V^{'(p)}_-(x)}{q + \mu_-^{(p)}}\right)
\end{equation}
and
\begin{eqnarray} \nonumber 
\tilde{P}(x,\ell,p|x_0) &=& \tilde{G}_\infty(x,p|x_0) \delta(\ell) \\   \label{eq:Pfull_int}
&+& \frac{1}{D} \left( V_+^{(p)}(x_0) V^{'(p)}_+(x) e^{-\ell \mu_+^{(p)}} 
+ V_-^{(p)}(x_0) V^{'(p)}_-(x) e^{-\ell\mu_-^{(p)}}\right).
\end{eqnarray}
Note that when the particle starts from one of the endpoints, the
first term vanishes, and one also has $V_{\pm}^{(p)}(0) =
v_{\pm}^{(p)}(0)$ and $V_{\pm}^{(p)}(L) = e^{\gamma L}
v_{\pm}^{(p)}(L)$.  While the Green's function $\tilde{G}_q(x,p|x_0)$
for partially reactive endpoints was known (and could be derived in a
direct way, see \ref{sec:Green}), the explicit form
(\ref{eq:Pfull_int}) of the Laplace-transformed full propagator
presents a new result.  These expressions in the no drift limit are
detailed in \ref{sec:no_drift}.

\subsection{Mean boundary local time}

Rewriting Eq. (\ref{eq:Vn_int}) explicitly as
\begin{equation} 
\overline{V_{\pm}^{'(p)}} = \frac{D \mu_{\pm}^{(p)}}{p} \biggl( v^{(p)}_{\pm}(0) + e^{-\gamma L} v^{(p)}_{\pm}(L) \biggr),
\end{equation}
we get from Eq. (\ref{eq:ellt_moments}) the Laplace transform of the
mean boundary local time:
\begin{eqnarray} \nonumber
\int\limits_0^\infty dt \, e^{-pt} \, \E_{x_0}\{ \ell_t\} 
&=& \frac{1}{p} \sum\limits_{\pm} \frac{v_{\pm}^{(p)}(0) + e^{-\gamma L} v_{\pm}^{(p)}(L)}{\mu_{\pm}^{(p)}} \, \\  \label{eq:Eell}
&\times&
\frac{e^{\gamma x_0}}{\sinh(\beta L)} \bigl(v_{\pm}^{(p)}(0) \sinh(\beta \bar{x}_0) + v_{\pm}^{(p)}(L) \sinh(\beta x_0)\bigr),
\end{eqnarray}
where the sum contains only two terms.  Note that higher-order moments
of $\ell_t$ can be written in a similar way.  Using the exact solution
(\ref{eq:Eell}) in the Laplace domain, one can analyze the short-time
and long-time behavior of the mean boundary local time.  Here we only
sketch the major points whereas the computational details are given in
\ref{sec:short}.

The long-time regime refers to the situation when a particle has
enough time to explore the whole confining domain a number of times,
i.e., $Dt \gg L^2$.  This regime corresponds to the limit $p\to 0$,
for which $\mu_+^{(p)} \approx 2\gamma \ctanh(\gamma L) + O(p)$ and
$\mu_-^{(p)} \approx p \frac{\tanh(\gamma L)}{2\gamma D} + O(p^2)$.
As the other ``ingredients'' in Eq. (\ref{eq:Eell}) converge to finite
limits as $p\to 0$ (and $\gamma \ne 0$), one deduces the long-time
asymptotic behavior:
\begin{equation}  \label{eq:Eell_t_long}
\E_{x_0}\{ \ell_t \} \simeq 2 \gamma \,\ctanh(\gamma L) Dt + O(1) \qquad (t\to\infty).
\end{equation}
Expectedly, the long-time behavior does not depend on the starting
point $x_0$.  In the no drift limit, $\gamma \to 0$, the constant in
front of the leading term approaches $2D/L$, yielding
\begin{equation} \label{eq:Eelltinf_nodrift}
\E_{x_0}\{ \ell_t \} \simeq \frac{2Dt}{L} + O(1) \qquad (t\to \infty) ,
\end{equation}
in agreement with the general long-time behavior for ordinary
diffusion in bounded domains \cite{Grebenkov19}.  The factor $\gamma\,
\ctanh(\gamma L)$ is an even monotonously increasing function:
whatever the sign of the drift, it tends to increase the mean boundary
local time in the long-time limit by keeping the particle closer to
the boundary and thus enhancing their encounters.

The behavior is totally different in the short-time limit.  First, the
initial position of the particle plays the crucial role.  Indeed, the
particle started in the bulk (i.e., $0 < x_0 < L$) needs some time for
traveling to the boundary in order to make the very first encounter
(i.e., to get $\ell_t > 0$).  As a consequence, the mean
$\E_{x_0}\{\ell_t\}$ vanishes exponentially fast as $t\to 0$ (see
\ref{sec:short} for details).  In contrast, when the particle starts
on the boundary, $\E_{x_0}\{\ell_t\}$ exhibits a power-law behavior.
We derived the short-time asymptotic formula (\ref{eq:Eell0_short})
for $\E_{0}\{\ell_t\}$ that can be written as
\begin{equation}
\E_{0}\{ \ell_t \} \approx \frac{F(-\gamma\sqrt{Dt})}{\gamma}  \qquad (t\to 0),
\end{equation}
with
\begin{equation}
F(x) = - \frac12 \erf(x) + x^2\, (1-\erf(x)) - \frac{|x| \, e^{-x^2}}{\sqrt{\pi}}  \,,
\end{equation}
where $\erf(x)$ is the error function.  By introducing the time scale
$t_\mu = 1/(D\gamma^2) = 4D/\mu^2$ associated to the drift, we can
distinguish the cases of positive and negative drifts, as well as the
limits of very short ($t \ll t_\mu$) and intermediate short ($t_\mu
\ll t \ll L^2/D$) times.

(i) For a positive drift (i.e., $\mu > 0$ and $\gamma < 0$), one has
$F(x) \simeq -2x/\sqrt{\pi} + x^2 + O(x^3)$ as $x = |\gamma|\sqrt{Dt}
\to 0$, and thus
\begin{equation}  \label{eq:Eell_short_pos}
\E_{0}\{ \ell_t \} \approx \frac{2\sqrt{Dt}}{\sqrt{\pi}} - \frac12 |\mu| t  \qquad (t \ll t_\mu, ~~ \mu > 0).
\end{equation}
The leading, $\sqrt{t}$-term remains unaffected by the drift, which
enters in the next-order correction.  In particular, one retrieves the
result from \cite{Grebenkov19} for restricted diffusion without
drift.  In turn, for intermediate short times $t\gg t_\mu$, one uses
$F(x) \approx 1/2$ as $x\to \infty$ to get
\begin{equation}  \label{eq:Eell_inter_pos}
\E_{0}\{ \ell_t \} \approx \frac{D}{2|\mu|}  \qquad (t_\mu \ll t \ll L^2/D, ~~ \mu > 0).
\end{equation}
In summary, a positive drift slightly diminishes the mean boundary
local time via the next-order correction in
Eq. (\ref{eq:Eell_short_pos}), and then leads to a constant value
controlled by $|\gamma|$, before reaching the long-time regime.  This
behavior agrees with the probabilistic picture, in which a positive
drift facilitates the departure of the particle from the left endpoint
$x_0$.  In a typical realization, the particle started from $x_0 = 0$
hits the left endpoint a number of times and then diffuses towards the
right endpoint.  The plateau in Eq. (\ref{eq:Eell_inter_pos})
corresponds, on average, to the passage from the left to the right
endpoint, during which the boundary local time does not change.

(ii) For a negative drift (i.e., $\mu < 0$ and $\gamma > 0$), the
situation is different.  At very short times, one has $F(x)
\simeq x^2 + O(x^3)$ as $x = - \gamma\sqrt{Dt} \to 0$ and thus
\begin{equation}  \label{eq:Eell_short_neg}
\E_{0}\{ \ell_t \} \approx \frac12 |\mu| t  \qquad (t \ll t_\mu, ~~ \mu < 0).
\end{equation}
Here, the linear slope, which is generally reminiscent of the
long-time behavior, is controlled by the drift, which keeps the
particle in the vicinity of the left endpoint.  For intermediate short
times, one has $F(x) \simeq 2x^2 + 1/2$ as $x\to -\infty$.  Neglecting
the constant $1/2$ in comparison to $2x^2$ in this limit, one gets
\begin{equation}  \label{eq:Eell_inter_neg}
\E_{0}\{ \ell_t \} \approx |\mu| t  \qquad (t_\mu \ll t\ll L^2/D, ~~ \mu < 0),
\end{equation}
i.e., the linear growth remains valid at intermediate times as well.
Finally, at long times, the linear scaling is retrieved again, with
the slope being controlled by the length of the interval (if the 
drift is not too strong).

\begin{figure}
\begin{center}
\includegraphics[width=100mm]{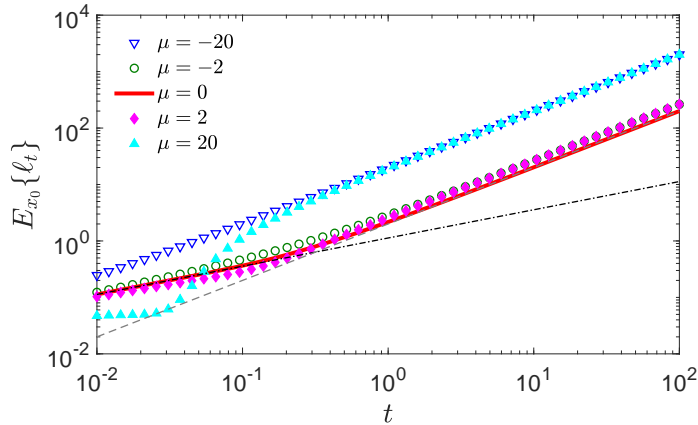} % Eell_t.eps}
\end{center}
\caption{
Mean boundary local time $\E_{x_0}\{\ell_t\}$ on the unit interval ($L
= 1$), with $D=1$, $x_0 = 0$, and five values of $\mu$, as shown in
the legend.  Dashed and dash-dotted lines illustrate respectively the
short-time relation (\ref{eq:Eell_short_pos}) with $\mu= 0$ and the
long-time relation (\ref{eq:Eelltinf_nodrift}) for the case without
drift.  $\E_{x_0}\{\ell_t\}$ was obtained via a numerical Laplace
transform inversion by a modified Talbot algorithm (see
\ref{sec:Talbot}). }
\label{fig:Eellt}
% A_localtime8_drift_Eell_t_fig;
\end{figure}

Figure \ref{fig:Eellt} illustrates the above picture.  The red solid
line presents the mean boundary local time $\E_{0}\{ \ell_t \}$ for
restricted diffusion without drift, which agrees well with both
long-time and short-time asymptotic relations
(\ref{eq:Eelltinf_nodrift}, \ref{eq:Eell_short_pos}).  Four other
curves plotted by symbols show $\E_{0}\{ \ell_t \}$ in the presence of
drift (with four different values of $\mu$).  At long times, the drift
shifts the curves upwards in the loglog plot due the prefactor $\gamma
\ctanh(\gamma L)$ in front of $t$ in Eq. (\ref{eq:Eell_t_long}).  The
effect is the same for both positive and negative drifts.
At short times, we retrieve the asymptotic behavior described above by
Eqs. (\ref{eq:Eell_short_pos} -- \ref{eq:Eell_inter_neg}) for positive
and negative drifts.  In particular, one clearly sees the plateau
region in the case $\mu = 20$.  In turn, this region is not present
for $\mu = 2$ because the time window of this regime, $t_\mu \ll t \ll
L^2/D$, is empty, given that $t_\mu = L^2/D = 1$.

\subsection{Distribution of the boundary local time}

To get the distribution of the boundary local time $\ell_t$, it is
enough to integrate the full propagator over $x$.  In fact, if
$P(\circ,\ell,t|x_0)$ denotes the probability density of $\ell_t$, its
Laplace transform reads
\begin{equation}  \fl  \label{eq:Pell_int}
\tilde{P}(\circ,\ell,p|x_0) = \tilde{S}_\infty(p|x_0) \delta(\ell)
+ \frac{1}{D} \left( V_+^{(p)}(x_0) \overline{V_+^{'(p)}} e^{-\ell \mu_+^{(p)}} +
V_-^{(p)}(x_0) \overline{V_-^{'(p)}} e^{-\ell \mu_-^{(p)}} \right),
\end{equation}
where
\begin{equation} 
\tilde{S}_\infty(p|x_0) = \frac{1}{p} \left(1 - \frac{\sinh(\beta \bar{x}_0) e^{\gamma x_0} 
+ \sinh(\beta x_0) e^{-\gamma \bar{x}_0}}{\sinh(\beta L)}\right)
\end{equation}

\begin{figure}
\begin{center}
\includegraphics[width=100mm]{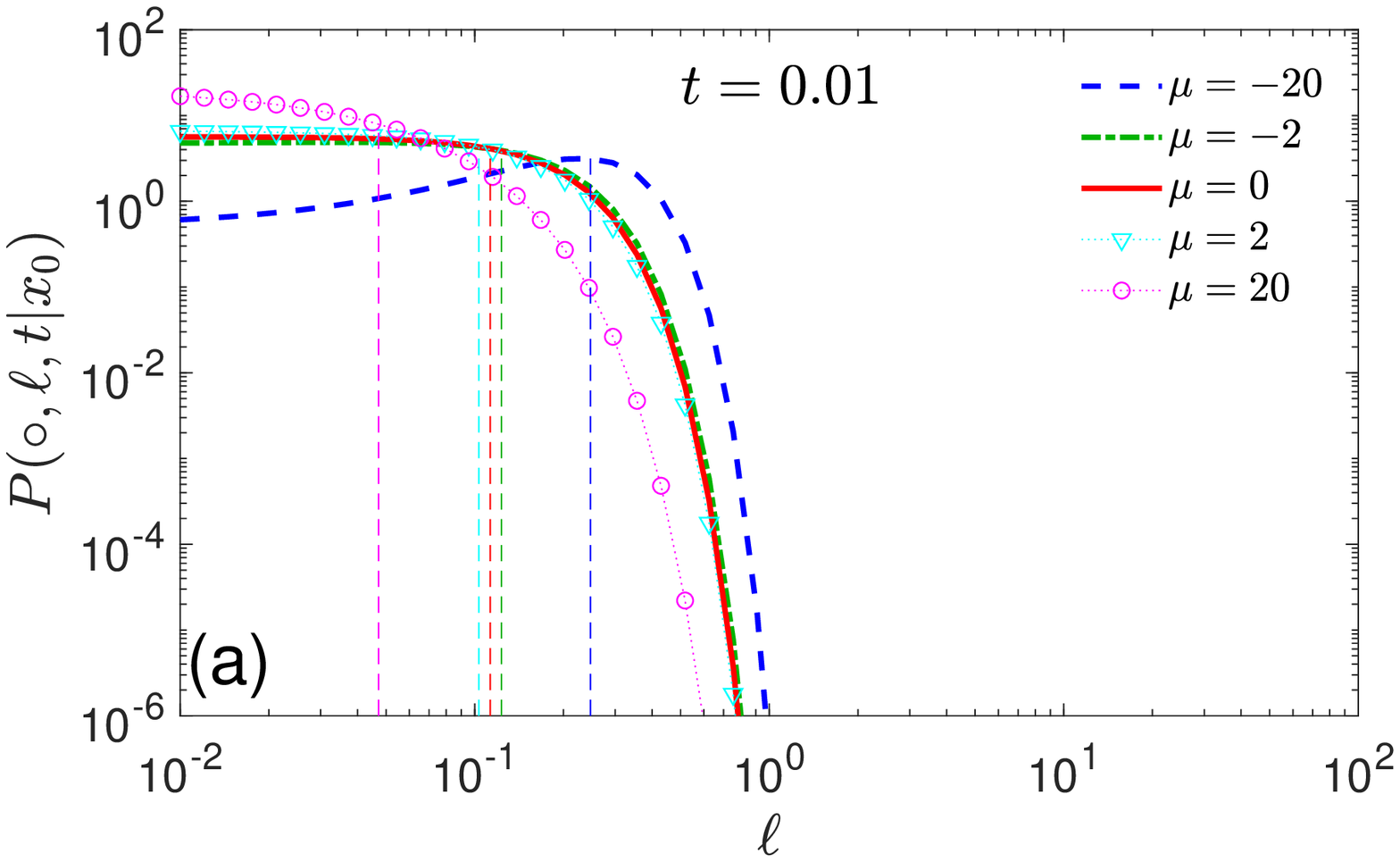} % Pellt_t001.eps}
\includegraphics[width=100mm]{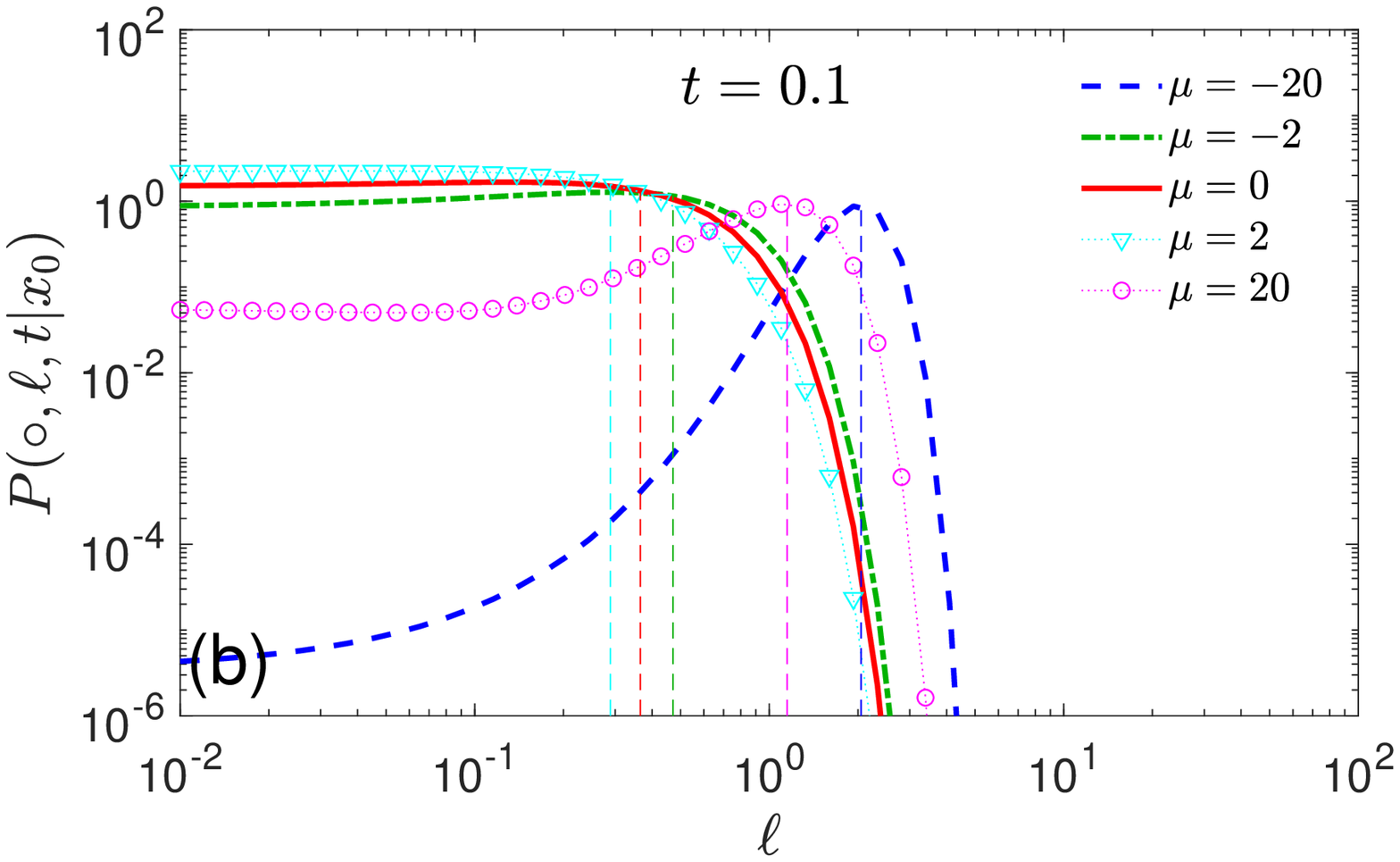} % Pellt_t01.eps}
\includegraphics[width=100mm]{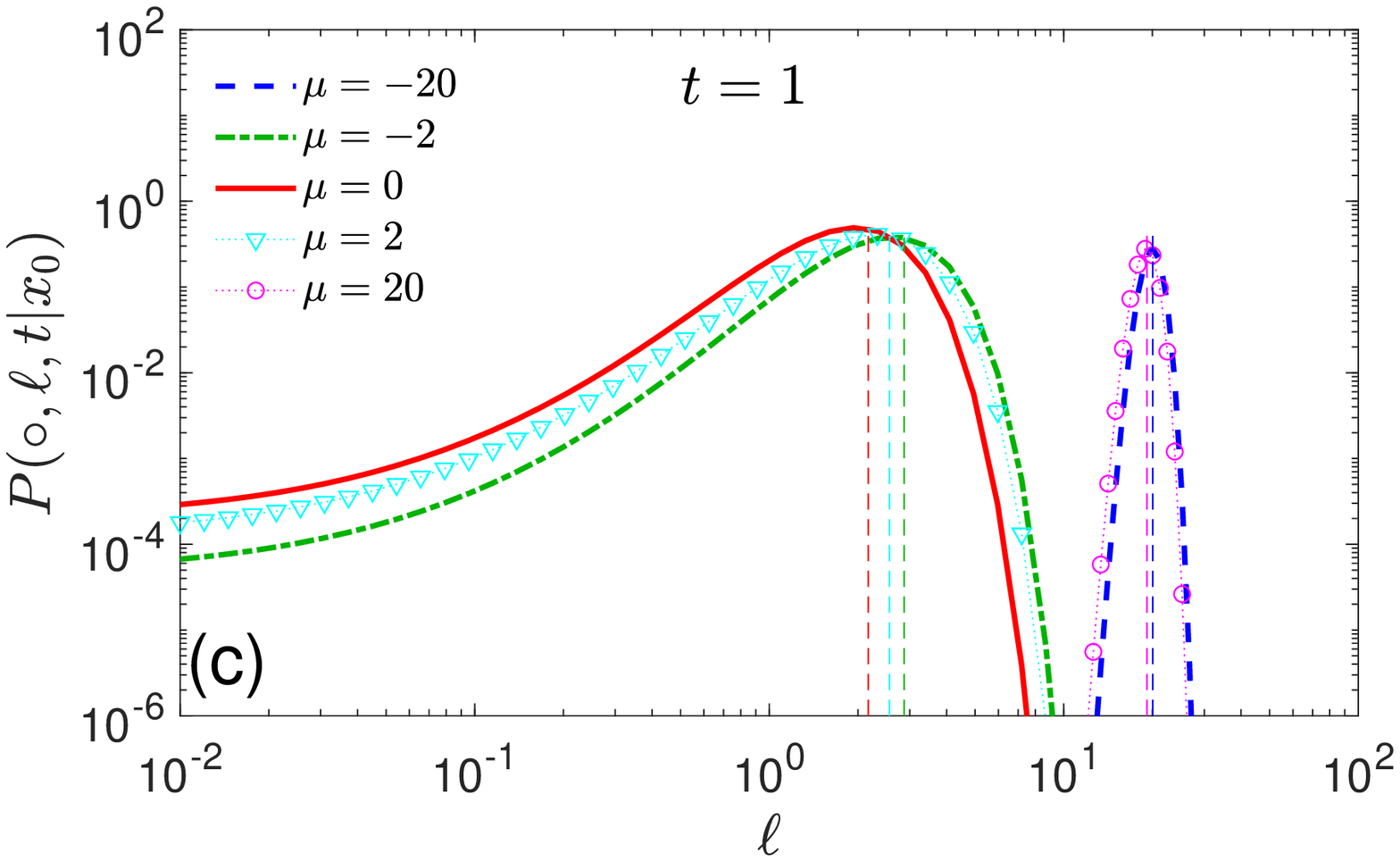} % Pellt_t1.eps}
\end{center}
\caption{
Probability density $P(\circ,\ell,t|x_0)$ of the boundary local time
$\ell_t$ on the unit interval ($L = 1$), with $D=1$, $x_0 = 0$, and
five values of $\mu$ (given in the legend) and three values of $t$:
{\bf (a)} $t = 0.01$, {\bf (b)} $t = 0.1$, and {\bf (c)} $t = 1$.
Vertical dashed lines indicate the corresponding mean boundary local
times $\E_0\{\ell_t\}$.  Both $P(\circ,\ell,t|x_0)$ and
$\E_0\{\ell_t\}$ were obtained via a numerical Laplace transform
inversion by a modified Talbot algorithm (see \ref{sec:Talbot}). }
\label{fig:Pell_t}
% A_localtime8_drift_Pell_t_fig2(0.01);
% A_localtime8_drift_Pell_t_fig2(0.1);
% A_localtime8_drift_Pell_t_fig2(1);
\end{figure}

For illustration purposes, we consider the situation when the particle
is already on the boundary, so that there is no contribution from the
first term in Eq. (\ref{eq:Pell_int}).  As the problem is invariant
under reflection with respect to the center of the interval (i.e.,
changing $x_0$ to $L-x_0$) along with changing $\mu$ to $-\mu$, one
can focus on the starting point $x_0 = 0$.  Figure \ref{fig:Pell_t}
illustrates the behavior of the probability density
$P(\circ,\ell,t|0)$ of the boundary local time $\ell_t$.  According to
Eq. (\ref{eq:Pell_int}), the Laplace transform of this function
reaches a constant as $\ell\to 0$, and rapidly vanishes at large
$\ell$.  These properties are preserved in time domain, even though
the value of the constant at small $\ell$ depends on time $t$.  As the
short-time limit corresponds to $p\to \infty$, the value
$\tilde{P}(\circ,0,p|0)$ weakly depends on the drift, and so does
$P(\circ,0,t|0)$, as one can see on the panel (a) for $t = 0.01$.
Expectedly, the distribution is shifted to the right (towards larger
$\ell$) by a negative drift which keeps the particle in the vicinity
of the left endpoint and thus facilitates encounters.  In contrast, a
positive drift has the opposite effect.  At an intermediate time $t =
0.1$ (panel (b)), the small-$\ell$ behavior of $P(\circ,\ell,t|0)$ is
more sensitive to the drift, in particular, the constant plateau at
$\ell\to 0$ is strongly attenuated by a strong negative drift ($\mu =
-20$) because small values of $\ell$ are highly unlikely.  A similar
trend is also visible for strong positive drift ($\mu = 20$) which
facilitates the passage to the right endpoint.  Finally, at moderately
long time $t = 1$ (panel (c)), there is almost no difference in the
effect of positive and negative drifts.  The distribution becomes
peaked around the mean value at large drifts $\mu = \pm 20$.

\begin{figure}
\begin{center}
\includegraphics[width=70mm]{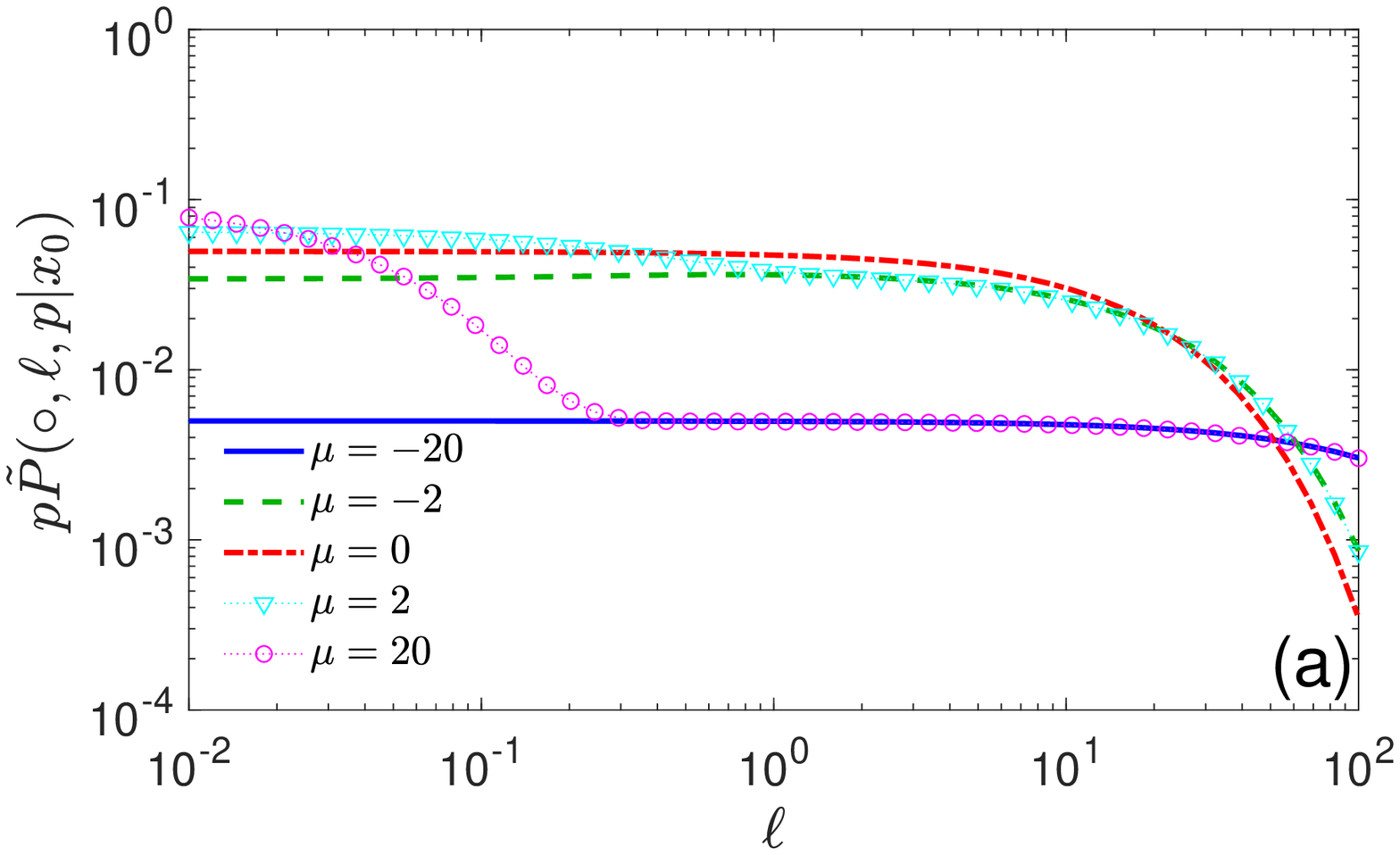} % Pell_p01.eps}   
\includegraphics[width=70mm]{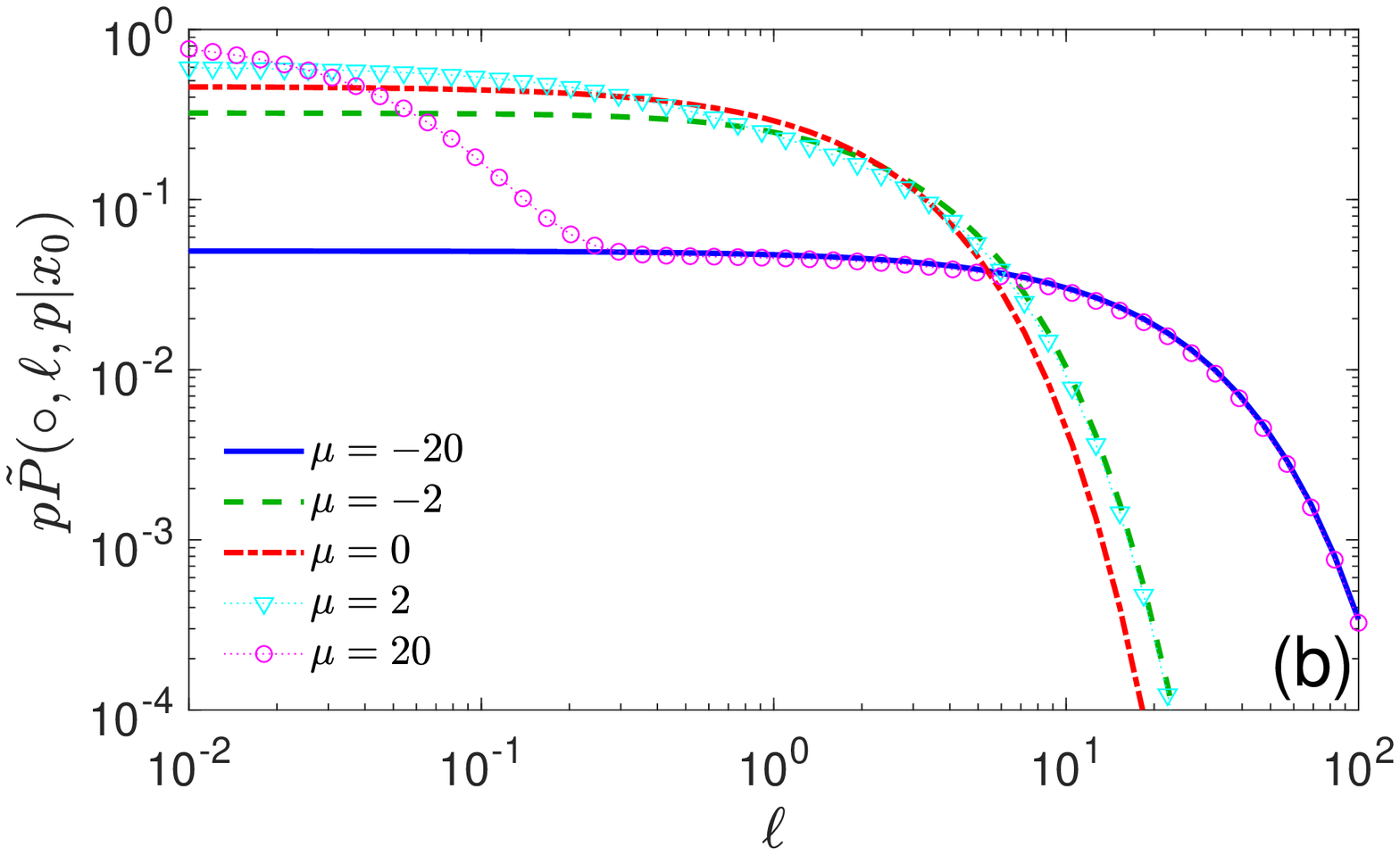} % Pell_p1.eps}
\end{center}
\caption{
Probability density $p\tilde{P}(\circ, \ell,p|x_0)$ of the boundary
local time $\ell_\tau$ (stopped at a random time $\tau$ exponentially
distributed with rate $p$) on the unit interval ($L = 1$), with $D=1$,
five values of $\mu$ (given in the legend), and two values of the rate
$p$: {\bf (a)} $p = 0.1$ and {\bf (b)} $p = 1$. }
\label{fig:Pell}
% A_localtime8_drift_Pell_fig3(0.1);
% A_localtime8_drift_Pell_fig3(1);
\end{figure}

To complete this section, we briefly discuss the probability density
$p\tilde{P}(\circ, \ell,p|x_0)$ of the boundary local time
$\ell_\tau$, which is stopped at a random time $\tau$ with an
exponential distribution: $\P\{\tau > t\} = e^{-pt}$.  Such a stopping
time can describe restricted diffusion in a reactive medium, in which
the particle can spontaneously disappear with the rate $p$;
alternatively, it can describe mortal walkers with a finite lifetime
\cite{Meerson15,Grebenkov17c}.  In this setting, we are interested in
the boundary local time at the death moment, i.e., how many times the
particle encountered the boundary during its lifetime.  According to
Eq. (\ref{eq:Pell_int}), the probability density of $\ell_\tau$ is
simply a bi-exponential function, whose decay rates are controlled by
the eigenvalues $\mu_\pm^{(p)}$.  In this setting,
Eq. (\ref{eq:Pell_int}) does not require an inverse Laplace transform
anymore and thus allows for a complete, fully explicit and elementary
study.  We skip this analysis and illustrate the behavior of the
probability density on Fig. \ref{fig:Pell}.  As previously, a negative
drift keeps the particle near the left endpoint and enhances its
encounters with that boundary, yielding higher probability for larger
values of $\ell_\tau$.  In turn, a strong positive drift results in
two constant levels of the probability density: a higher plateau at
small $\ell$, which corresponds to encounters with the right endpoint,
and a lower plateau at larger $\ell$, which represents the
contribution of encounters with the left endpoint (as in the case of
the negative drift).  In fact, when the particle leaves the left
endpoint, it spends a considerable part of its lifetime to diffuse
towards the right endpoint, during which the boundary local time does
not change.  As a consequence, one gets small $\ell_\tau$ for such
realizations.  We note that the qualitative behavior for two
considered reaction rates $p = 0.1$ and $p = 1$ is very similar,
except that the decay at large $\ell$ is faster for $p = 1$ because
the particles with the shorter lifetime get on average smaller
$\ell_\tau$.

\section{Discussion}
\label{sec:discussion}

In this paper, we extended the encounter-based approach developed in
\cite{Grebenkov20} to restricted diffusion with a gradient drift.
This approach relies on the Langevin (or Skorokhod) stochastic
differential equation that describes simultaneously the position of a
particle and its boundary local time.  Quite naturally, their joint
probability density $P(\x,\ell,t|\x_0)$, that we called the full
propagator, appears to be the fundamental characteristics of
restricted diffusion.  This quantity is independent of surface
reactions and thus properly describes the dynamics alone.  In turn,
the conventional propagator, which accounts for surface reactions
through the Robin boundary condition, can then be deduced as the
Laplace transform of $P(\x,\ell,t|\x_0)$ with respect to the boundary
local time $\ell$.  This relation reflects the Bernoulli character of
the surface reaction mechanism when the particle attempts to react at
each encounter with equal probabilities and these trials are
independent from each other.  Most importantly, other surface reaction
mechanisms can be implemented in a very similar way by replacing the
exponential stopping condition.  The disentanglement of the diffusive
dynamics from surface reactions presents thus one of the major
advantages of the encounter-based approach.
%Note that this idea was earlier
%exploited by Szabo {\it et al.} in the particular case of a finite
%number of point-like target sites in the bulk \cite{Szabo84}.

\begin{figure}
\begin{center}
\includegraphics[width=70mm]{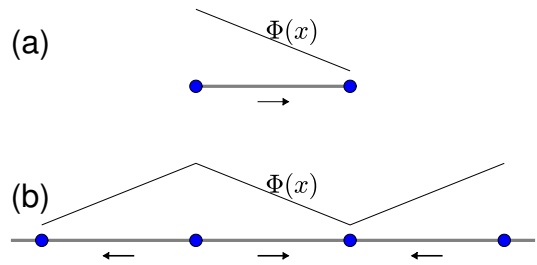} % scheme1.eps}   
\end{center}
\caption{
{\bf (a)} An interval of length $L$ with two reflecting endpoints and
a linearly decreasing potential $\Phi(x)$ that describes a positive
drift; {\bf (b)} An infinite line with equally spaced discrete points
$\{n L\}$ and a triangular periodic potential $\Phi(x)$ that describes
alternating positive and negative drifts.}
\label{fig:scheme}
% A_localtime8_drift_scheme();
\end{figure}

In order to derive a spectral decomposition for the
Laplace-transformed full propagator, we introduced an appropriate
Dirichlet-to-Neumann operator based on the symmetrized version of the
Fokker-Planck operator.  We illustrated the advantages of this
approach in the case of restricted diffusion on an interval with a
constant drift.  In particular, we analyzed the distribution of the
boundary local time and showed the effects of both positive and
negative drifts at short and long times.  We note that restricted
diffusion on the interval $(0,L)$ with reflecting endpoints is
equivalent to that on an infinite line and on a circle.  As a
consequence, the obtained exact formula for the distribution of the
boundary local time can also describe (i) the boundary local time on
discrete points $\{ nL\}_{n\in \Z}$ equally spaced on an infinite line
with a periodic triangular potential (see Fig. \ref{fig:scheme}), and
(ii) the boundary local time on an even number of equally spaced points
on a circle.

Throughout the manuscript, the whole boundary was supposed to be
reactive.  In many applications, however, one deals with a reactive
region $\RR \subset \pa$ on otherwise inert impermeable boundary, and
aims at finding the distribution of first-reaction times on that
region.  In this case, the Robin boundary condition
(\ref{eq:BC_forward}) is replaced by mixed Robin-Neumann boundary
condition 
\begin{equation}
j_q(\s,t|\x_0) = \left\{ \begin{array}{ll} \kappa G_q(\s,t|\x_0)  & (\s\in\RR), \\  0 & (\s\in\pa\backslash\RR). \\ \end{array} \right.
\end{equation}
As discussed in \cite{Grebenkov20c}, an extension of the
encounter-based approach to this setting is straightforward.  In fact,
one focuses on the boundary local time on the reactive region $\RR$,
which is obtained by substituting $\pa$ by $\RR$ in
Eq. (\ref{eq:ellt_def}).  The former derivations remain valid, if the
definition (\ref{eq:DtN}) of the Dirichlet-to-Neumann operator $\M_p$
is generalized as
\begin{equation} \fl  \label{eq:DtN2}
\M_p f = \bigl(\partial_n w + \phi(\x) w\bigr)|_{\RR},  \quad \textrm{with} ~~ 
\left\{ \begin{array}{ll} (p - \bar{\L}_{\x}) w = 0 & (\x\in \Omega), \\  w = f & (\x\in \RR),  \\ 
\partial_n w + \phi(\x) w = 0 & (\x\in\pa\backslash \RR).
\end{array} \right.
\end{equation}
In other words, the operator $\M_p$ associates to a function $f$ on
the reactive region $\RR$ another function on $\RR$, keeping the
reflecting boundary condition for the solution $w$ on the remaining
inert region $\pa\backslash \RR$.  Examples of this extension for
ordinary diffusion were given in \cite{Grebenkov20c}.  As the reactive
region does not need to be connected, the problem of multiple reactive
sites can be treated in this framework.  In addition, one can include
an absorbing region $\AA \subset \pa$ that kills the particle upon the
first encounter.  In this way, one can investigate surface reactions
on $\RR$ for a sub-population of particles that avoid hitting $\AA$.
An implementation of this setting consists in adding the boundary
condition $w = 0$ on $\AA$ in the definition (\ref{eq:DtN2}) of the
Dirichlet-to-Neumann operator (see further discussions in
\cite{Grebenkov20b}).  To get more subtle insights onto reactions on
different reactive sites (e.g., competition between them), one can
introduce the individual boundary local time on each site and look at
their joint distribution \cite{Grebenkov20b}.  However, appropriate
spectral decompositions in this setting are yet unknown.

The proposed encounter-based approach can be further developed in
different directions.  On one hand, one can dwell on a rigorous
proof of the spectral decompositions, on mathematical conditions for
the boundary smoothness, on the spectral properties of the introduced
extension of the Dirichlet-to-Neumann operator $\M_p$, and on their
probabilistic interpretations.  In particular, we assumed that $\M_p$
is a self-adjoint operator with a discrete spectrum, without
discussing eventual limitations on the drift $\mmu(\x)$.  Moreover,
one can further extend the present approach to unbounded domains with
a compact boundary (e.g., the exterior of a ball).  While such an
extension was already realized for diffusion without drift (see, e.g.,
\cite{Grebenkov21}), some limitations on the drift $\mmu(\x)$ can be
expected to ensure the discrete spectrum of the Dirichlet-to-Neumann
operator.  Yet another mathematical direction is related to the
asymptotic analysis of the narrow-escape problem when only a small
fraction of the boundary is reactive.  On the other hand, one can
further explore the advantages of the encounter-based approach for
various physical settings.  For instance, one can consider the effect
of local potentials near the boundary (such as, e.g., electrostatic
repulsion or attractive interactions) onto the statistics of boundary
encounters.  One can consider other surface reaction mechanisms
\cite{Grebenkov20} and investigate how the drift may affect them.
To some extent, such an approach can provide a microscopic model for
so-called intermittent diffusions, in which a particle alternates
diffusions in the bulk and on the surface
\cite{Levitz08,Chechkin09,Benichou10,Benichou11,Chechkin11,Chechkin12,Rupprecht12a,Rupprecht12b,Berezhkovskii15,Berezhkovskii17}.
Similarly, this approach may also help for studying diffusion
with reversible binding to the boundary, the so-called sticky boundary
condition, when the particle stays on the boundary for a random
waiting time and then unbinds to resume its bulk diffusion.  The
effects of this reversible binding onto reaction rates and
first-passage times were thoroughly investigated for ordinary
diffusion (see
\cite{Agmon90,Prustel13,Grebenkov17f,Lawley19,Reva21} and references
therein), as well as for more sophisticated processes such as
diffusion with stochastic resetting
\cite{Evans11,Chechkin18,Evans19,Evans20} or velocity jump processes
\cite{Gopalakrishnan11,Zelinski12,Mulder12,Angelani17,Bressloff19}.
Since binding to the boundary can be understood as a reaction event on
that boundary, the binding time can be described as the first moment
when the boundary local time exceeds an exponentially distributed
threshold, as in Eq. (\ref{eq:stopping}).  Changing the probability
law for the random threshold allows one to incorporate more
sophisticated binding mechanisms (see \cite{Grebenkov20} for details).
Curiously, the unbinding event implies resetting of the boundary local
time.

Numerous advantages of the encounter-based approach urge for its
extensions to even more general diffusion processes, in particular,
with general space-dependent drift and volatility matrix.  Such an extension
would have to overcome two major difficulties.

(i) When the diffusivity varies in the bulk, the
reactivity parameter $q = \kappa/D$ becomes space-dependent as well.
As a consequence, the probability of reaction at each encounter
depends on the position of that encounter.  This dependence would thus
prohibit using the Laplace transform relation (\ref{eq:Gq_P}) between
the full propagator and the conventional propagator.  Following
\cite{Papanicolaou90}, one would thus need to introduce a functional
of $q(\X_t)$ to account for the cumulative effect of multiple reaction
attempts (see also discussion in \cite{Grebenkov20b}).  Note that a
similar difficulty appears in the case of ordinary diffusion towards a
heterogeneously reactive boundary, in which case the space-dependent
reactivity $\kappa$ renders $q$ to be space-dependent as well
\cite{Grebenkov19}.  

(ii) Even if the diffusivity remains constant, an extension to a
general drift becomes challenging.  In fact, the Fokker-Planck
operator can be reduced to a self-adjoint form with real eigenvalues
only in the case of a gradient drift \cite{Nelson58}.  Even though an
appropriate Dirichlet-to-Neumann operator can potentially be
introduced in a more general case, such an operator is likely to be
non-self-adjoint that would raise considerable challenges in the
spectral analysis.  Moreover, the numerical computation of the related
spectral properties may also be difficult.  Nevertheless, further
mathematical works in this direction are expected to shed light onto
more general diffusion-influenced reactions.

More generally, one can go beyond the framework of stochastic
differential equations with Gaussian noises.  In the simplest case of
random walks on a lattice or a graph, the boundary local time is the
number of visits of a given subset of vertices, which can be
interpreted as partially reactive sites or localized traps.  Szabo
{\it et al.}  showed how to express the related propagator in terms of
the propagator without trapping \cite{Szabo84}.  This concept was
recently extended to more general random walks \cite{Guerin21}.
Another interesting extension concerns diffusion processes with
stochastic resetting \cite{Evans11,Chechkin18,Evans19,Evans20}, which
can be implemented either for the position, or for the boundary local
time, or for both processes.  The renewal character of such resettings
may allow getting rather explicit results in the Laplace domain.
Moreover, the main idea of the encounter-based approach can be
potentially applied to other examples of stochastic dynamics such as,
e.g., velocity jump processes (including run-and-tumble motion)
\cite{Gopalakrishnan11,Zelinski12,Mulder12,Angelani17,Bressloff19} and
continuous-time random walks (including even L\'evy flights).
However, one would need to properly define the boundary local time (or
its analog), as well as an appropriate governing operator (such as the
Dirichlet-to-Neumann operator), whose spectral properties would
determine the full propagator.  Feasibility, mathematical rigor and
practical value of such extensions are still open.

\ack

The author acknowledges the Alexander von Humboldt Foundation for
support within a Bessel Prize award.

\appendix

\section{Auxiliary relation}
\label{sec:auxil}

In this Appendix, we obtain Eq. (\ref{eq:Vn_int}) by extending the
derivation from \cite{Grebenkov19}.

First, the integral of Eq. (\ref{eq:Gq_fFP}) over $\x\in\Omega$ yields
\begin{eqnarray*} 
p \int\limits_{\Omega} d\x \, \tilde{G}_\infty(\x,p|\x_0) &=& 1 + \int\limits_{\Omega}d\x \, (D \Delta - (\nabla \mmu)) \tilde{G}_\infty(\x,p|\x_0) \\
&=& 1 + \int\limits_{\pa} d\x \, D \partial_n \tilde{G}_\infty(\x,p|\x_0).
\end{eqnarray*}
Taking the normal derivative with respect to $\x_0$ at $\x_0 =
\s_0\in\pa$ and multiplying by $-D/p$, we get
\begin{equation}  \label{eq:auxil33}
\int\limits_{\Omega} d\x \, \tilde{j}'_\infty(\s_0,p|\x) 
= -\frac{D}{p} \int\limits_{\pa} d\x \, D (\partial_{n_0} \partial_n \tilde{G}_\infty(\x,p|\x_0))|_{\x_0=\s_0} .
\end{equation}
Our intention is to exchange the order of normal derivatives in the
right-hand side in order to get $\tilde{j}'_{\infty}(\s_0,p|\x)$.
However, as this function is singular in the vicinity of $\x\approx
\s_0$, one cannot simply exchange the order.  To proceed, we use
Eq. (\ref{eq:Gq_gq}) and evaluate the following derivatives:
\begin{eqnarray*} \fl
\partial_{n_0} \partial_{n} \tilde{G}_\infty(\x,p|\x_0) &=& \partial_{n_0} \partial_{n} 
e^{-\frac12\Phi(\x)+\frac12 \Phi(\x_0)} \tilde{g}_\infty(\x,p|\x_0) \\ \fl
&=& e^{-\frac12\Phi(\x)+\frac12 \Phi(\x_0)} \biggl( - \phi(\x) \partial_{n_0} \tilde{g}_\infty(\x,p|\x_0) + 
\partial_{n_0} \partial_{n} \tilde{g}_\infty(\x,p|\x_0)\biggr), \\ \fl
\partial_{n} \partial_{n_0} \tilde{G}_\infty(\x,p|\x_0) &=& \partial_{n} \partial_{n_0} 
e^{-\frac12\Phi(\x)+\frac12 \Phi(\x_0)} \tilde{g}_\infty(\x,p|\x_0) \\ \fl
&=& e^{-\frac12\Phi(\x)+\frac12 \Phi(\x_0)} \biggl( \phi(\x_0) \partial_{n} \tilde{g}_\infty(\x,p|\x_0) + 
\partial_{n} \partial_{n_0} \tilde{g}_\infty(\x,p|\x_0)\biggr).
\end{eqnarray*}
Since the function $\tilde{g}_\infty(\x,p|\x_0)$ is symmetric, one can
exchange the order of normal derivatives.  In addition, one has
\begin{eqnarray*} 
\tilde{j}_{\infty}(\x,p|\x_0) &=& - D \partial_n \biggl(e^{-\frac12\Phi(\x)+\frac12 \Phi(\x_0)} \tilde{g}_\infty(\x,p|\x_0)\biggr) \\  
&=& \phi(\x) D \tilde{G}_\infty(\x,p|\x_0) - D e^{-\frac12\Phi(\x)+\frac12 \Phi(\x_0)} \partial_n \tilde{g}_\infty(\x,p|\x_0), \\
\tilde{j}'_{\infty}(\x_0,p|\x) &=& - D \partial_{n_0} \biggl(e^{-\frac12\Phi(\x)+\frac12 \Phi(\x_0)} \tilde{g}_\infty(\x,p|\x_0)\biggr) \\  
&=& -\phi(\x_0) D \tilde{G}_\infty(\x,p|\x_0) - D e^{-\frac12\Phi(\x)+\frac12 \Phi(\x_0)} \partial_{n_0} \tilde{g}_\infty(\x,p|\x_0) .
\end{eqnarray*}
Combining these relations and noting that
$\tilde{j}_{\infty}(\x,p|\x_0) = \tilde{j}'_{\infty}(\x_0,p|\x) =
\delta(\x-\x_0)$ when both $\x$ and $\x_0$ belong to the boundary
$\pa$, we get
\begin{equation}
\partial_{n_0} \partial_{n} \tilde{G}_\infty(\x,p|\x_0) - \partial_{n} \partial_{n_0} \tilde{G}_\infty(\x,p|\x_0)
= \frac{\phi(\x) + \phi(\x_0)}{D} \delta(\x - \x_0).
\end{equation}
This relation allows us to rewrite Eq. (\ref{eq:auxil33}) as
\begin{equation}   \label{eq:auxil34}
\int\limits_{\Omega} d\x \, \tilde{j}'_\infty(\s_0,p|\x) 
= \frac{D}{p} \int\limits_{\pa} d\x \, (\partial_{n} \tilde{j}'_\infty(\s_0,p|\x)) - \frac{2D \phi(\s_0)}{p} .
\end{equation}

We multiply both sides of this equation by $e^{-\frac12\Phi(\s_0)}
f(\s_0)$ with a suitable function $f(\s_0)$ and integrate over $\s_0
\in \pa$:
\begin{eqnarray}  \nonumber
&& \int\limits_{\pa} d\s_0 \, e^{-\frac12 \Phi(\s_0)} f(\s_0) \int\limits_{\Omega} d\x \, \tilde{j}'_\infty(\s_0,p|\x)  \\   \label{eq:auxil21}
&&= \frac{D}{p} \int\limits_{\pa} d\s_0 e^{-\frac12 \Phi(\s_0)} f(\s_0)
 \left\{ \int\limits_{\pa} d\s \, (\partial_n \tilde{j}'_\infty(\s_0,p|\s)) - 2\phi(\s_0)\right\}.
\end{eqnarray}
Our aim is to show that the right-hand side can be represented via the
Dirichlet-to-Neumann operator.  For this purpose, we note that the
solution of the equation $(p-\bar{\L}) u = 0$ with Dirichlet boundary
condition $u = f$ on $\pa$ can be written as
\begin{eqnarray*} 
u(\x) &=& \int\limits_{\Omega} d\x_0 \biggl[u(\x_0) (p - \bar{\L}_{\x_0}) \tilde{g}_\infty(\x,p|\x_0) 
- \tilde{g}_\infty(\x,p|\x_0) (p - \bar{\L}_{\x_0}) u(\x_0) \biggr]  \\
&=& \int\limits_{\pa} d\x_0 u(\x_0) (-D\partial_{n_0} \tilde{g}_\infty(\x,p|\x_0)) \\
&=& \int\limits_{\pa} d\x_0 u(\x_0) (-D\partial_{n_0} e^{\frac12\Phi(x) - \frac12 \Phi(\x_0)} \tilde{G}_\infty(\x,p|\x_0)) \\
&=& \int\limits_{\pa} d\s_0 f(\s_0) e^{\frac12\Phi(\x) - \frac12 \Phi(\s_0)}  \tilde{j}'_\infty(\s_0,p|\x).
\end{eqnarray*}
As a consequence, the action of the Dirichlet-to-Neumann operator is
\begin{eqnarray*} \fl 
&& (\M_p f)(\s) = \phi(s) f(\s) + (\partial_n u)_{\x=\s} \\ \fl
&& = \phi(s) f(\s) + \int\limits_{\pa} d\s_0 f(\s_0) e^{ - \frac12 \Phi(\s_0)} 
\partial_n \bigl(e^{\frac12\Phi(\x)}  \tilde{j}'_\infty(\s_0,p|\x)\bigr)|_{\x=\s} \\ \fl
&& = \phi(\s) f(\s) + \int\limits_{\pa} d\s_0 f(\s_0) e^{ - \frac12 \Phi(\s_0)} 
e^{\frac12\Phi(\s)} \biggl[\phi(\s) \underbrace{\tilde{j}'_\infty(\s_0,p|\s)}_{=\delta(\s-\s_0)} 
+ \bigl(\partial_n \tilde{j}'_\infty(\s_0,p|\x)\bigr)|_{\x=\s}\biggr] \\ \fl
&& = 2\phi(\s) f(\s) + \int\limits_{\pa} d\s_0 f(\s_0) e^{ - \frac12 \Phi(\s_0)} 
e^{\frac12\Phi(\s)} \bigl(\partial_n \tilde{j}'_\infty(\s_0,p|\x)\bigr)|_{\x=\s}.
\end{eqnarray*}
This relation allows us to represent the integral over $\s_0$ in the
right-hand side of Eq. (\ref{eq:auxil21}) in terms of $\M_p$:
\begin{eqnarray*}  \nonumber
&& \int\limits_{\pa} d\s_0 \, e^{-\frac12 \Phi(\s_0)} f(\s_0) \int\limits_{\Omega} d\x \, \tilde{j}'_\infty(\s_0,p|\x)  \\    
&&= \frac{D}{p}\Biggl\{ \int\limits_{\pa} d\s \, e^{-\frac12\Phi(\s)} \, \underbrace{e^{\frac12\Phi(\s)} 
\int\limits_{\pa} d\s_0 e^{-\frac12 \Phi(\s_0)} f(\s_0) (\partial_n \tilde{j}'_\infty(\s_0,p|\s))}_{
= (\M_p f)(\s) - 2\phi(\s) f(\s)} \\
&& - 2\int\limits_{\pa} d\s_0 e^{-\frac12 \Phi(\s_0)} f(\s_0) \phi(\s_0)\Biggr\} ,
\end{eqnarray*}
i.e.,
\begin{equation}
\int\limits_{\pa} d\s_0 \, e^{-\frac12 \Phi(\s_0)} f(\s_0) \int\limits_{\Omega} d\x \, \tilde{j}'_\infty(\s_0,p|\x)
= \frac{D}{p} \int\limits_{\pa} d\s \, e^{-\frac12\Phi(\s)} (\M_p f)(\s).
\end{equation}
This relation holds for any suitable function $f(\s)$ on the boundary.
Setting $f(\s) = v_n^{(p)}(\s)$ to be an eigenfunction of $\M_p$, one
gets immediately
\begin{eqnarray}  \nonumber
\overline{V_n^{'(p)}} &=& 
\int\limits_{\Omega} d\x \, \underbrace{\int\limits_{\pa} d\s_0 \, e^{-\frac12 \Phi(\s_0)} v_n^{(p)}(\s_0) 
 \tilde{j}'_\infty(\s_0,p|\x)}_{=V_n^{'(p)}(\x)} \\  %\label{eq:Vn_int} 
&=& \frac{D}{p} \mu_n^{(p)} \int\limits_{\pa} d\s \, e^{-\frac12\Phi(\s)} v_n^{(p)}(\s).
\end{eqnarray}

\section{Green's functions on the interval}
\label{sec:Green}

Even though many Green's functions on the interval are known (see,
e.g., \cite{Thambynayagam}), we provide here the main steps of the
derivation for the Fokker-Planck operator $\L = D\partial_x^2 - \mu
\partial_x$ for completeness.

We start with the most general case of Robin boundary conditions at
$0$ and $L$, with two different reactivities $q_1$ and $q_2$.  A
general solution of the equation $(p-\L)u = 0$ reads $u(x) = c_+
e^{\alpha_+ x} + c_- e^{\alpha_- x}$, with
\begin{equation}
\alpha_{\pm} = \frac{\mu}{2D} \pm \sqrt{\frac{p}{D} + \frac{\mu^2}{4D^2}} \,,
\end{equation}
and two arbitrary coefficients $c_\pm$.  As usual, one searches for
the solution of the equation (\ref{eq:Gq_fFP}) in the form
\begin{equation} \fl
\tilde{G}_{q_1,q_2}(x,p|x_0) = \left\{ \begin{array}{ll}  A \bigl[(1-h_1 \alpha_-) e^{\alpha_+ x} - (1-h_1\alpha_+) e^{\alpha_- x}\bigr] & (0 < x < x_0), \\
B \bigl[(1+h_2 \alpha_-) e^{\alpha_+ (x-L)} - (1 + h_2 \alpha_+) e^{\alpha_- (x-L)}\bigr] & (x_0 < x < L), \\  \end{array} \right.
\end{equation}
that satisfies the boundary condition (\ref{eq:Gq_fFP_BC}) at both
endpoints:
\begin{eqnarray}
\bigl(-\partial_x + q_1 + \mu/D \bigr)  \tilde{G}_{q_1,q_2}(x,p|x_0) &=& 0  \quad (x = 0) ,\\
\bigl(\partial_x + q_2 - \mu/D \bigr)  \tilde{G}_{q_1,q_2}(x,p|x_0) &=& 0  \quad (x = L) ,
\end{eqnarray}
where we set $h_1 = 1/(q_1 + \mu/D)$ and $h_2 = 1/(q_2 - \mu/D)$.  The
unknown coefficients $A$ and $B$ are then determined by requiring the
continuity of $\tilde{G}_{q_1,q_2}(x,p|x_0)$ at $x= x_0$ and the drop
of the derivative by $-1/D$ (i.e., $[\partial_x
\tilde{G}_{q_1,q_2}(x,p|x_0)]_{x=x_0+\epsilon} - [\partial_x
\tilde{G}_{q_1,q_2}(x,p|x_0)]_{x=x_0-\epsilon} \to - 1/D$ as $\epsilon\to 0$):
\begin{eqnarray} \fl
A &=& \frac{e^{\gamma x_0} \bigl[(1-h_2\gamma) \sinh(\beta \bar{x}_0) + \beta h_2 \cosh(\beta \bar{x}_0)\bigr]}
{2\beta D \bigl[\sinh(\beta L) (1 + \gamma(h_1-h_2) + (\beta^2-\gamma^2) h_1h_2) + \cosh(\beta L) \beta (h_1+h_2)\bigr]} \,, \\ \fl
B &=& - \frac{e^{\gamma (x_0-L)} \bigl[(1+h_1\gamma) \sinh(\beta x_0) + \beta h_1 \cosh(\beta x_0)\bigr]}
{2\beta D \bigl[\sinh(\beta L) (1 + \gamma(h_1-h_2) + (\beta^2-\gamma^2) h_1h_2) + \cosh(\beta L) \beta (h_1+h_2)\bigr]} \,,
\end{eqnarray}  % checked in interval_drift_new.mw
where we used the former notations $\gamma = -\mu/(2D)$, $\beta =
\sqrt{p/D + \gamma^2}$, and $\bar{x}_0 = L-x_0$.  One can also write
\begin{equation*} \fl
\tilde{G}_{q_1,q_2}(x,p|x_0) = \left\{ \begin{array}{ll}  2A e^{-\gamma x} \bigl[(1+h_1 \gamma)\sinh(\beta x) + h_1 \beta \cosh(\beta x) \bigr] & (0 < x < x_0), \\
-2B e^{-\gamma(x-L)} \bigl[(1-h_2 \gamma) \sinh(\beta \bar{x}) + h_2
\beta \cosh(\beta \bar{x})\bigr] & (x_0 < x < L), \\ \end{array} \right.
\end{equation*}
with $\bar{x} = L - x$.  Setting
\begin{eqnarray} \fl
U(x) & =& (1+h_1 \gamma) \sinh(\beta x) + \beta h_1 \cosh(\beta x) , \\ \fl
V(x) & =& (1-h_2 \gamma) \sinh(\beta (L-x)) + \beta h_2 \cosh(\beta (L-x)), \\ \fl
C &= & \frac{1}{\beta D \bigl[\sinh(\beta L) (1 + \gamma(h_1-h_2) + (\beta^2-\gamma^2) h_1h_2) + \cosh(\beta L) \beta (h_1+h_2)\bigr]} ,
\end{eqnarray}
one can finally get
\begin{equation} 
\tilde{G}_{q_1,q_2}(x,p|x_0) = C e^{-\gamma (x-x_0)} \left\{ \begin{array}{ll}  U(x) V(x_0) & (0 < x < x_0), \\
U(x_0) V(x) & (x_0 < x < L). \\  \end{array} \right.
\end{equation}

In the particular case of equal reactivities, $q_1 = q_2 = q$, one
has
\begin{eqnarray*}
U(x) &=& \frac{\hat{U}(x)}{q-2\gamma},  \qquad \hat{U}(x) = (q-\gamma)\sinh(\beta x) + \beta \cosh(\beta x) ,\\
V(x) &=& \frac{\hat{V}(x)}{q+2\gamma},  \qquad \hat{V}(x) = (q+\gamma)\sinh(\beta (L-x)) + \beta \cosh(\beta (L-x)) ,\\
C &=& \frac{\hat{C}}{q^2 - 4\gamma^2} , \qquad \hat{C} = 
\frac{1}{\beta D \sinh(\beta L) \bigl( q^2 + 2\beta \ctanh(\beta L)q + \beta^2-\gamma^2 \bigr)} .
\end{eqnarray*}
More explicitly, one gets
\begin{eqnarray*} \fl
\tilde{G}_{q}(x,p|x_0) &=& e^{-\gamma (x-x_0)} \frac{\sinh(\beta \bar{x}_0)\sinh(\beta x)}{\beta D \sinh(\beta L)} \\
\fl 
&\times &
\frac{q^2 + q \beta (\ctanh(\beta \bar{x}_0) + \ctanh(\beta x)) - (\gamma - \beta \ctanh(\beta x))
 (\gamma + \beta \ctanh(\beta \bar{x}_0))}{q^2 + 2\beta \ctanh(\beta L)q + \beta^2 - \gamma^2}
\end{eqnarray*}
for $0 < x < x_0$, and
\begin{eqnarray*} \fl
\tilde{G}_{q}(x,p|x_0) &=& e^{-\gamma (x-x_0)} \frac{\sinh(\beta \bar{x})\sinh(\beta x_0)}{\beta D \sinh(\beta L)} \\
\fl 
&\times &
\frac{q^2 + q \beta (\ctanh(\beta \bar{x}) + \ctanh(\beta x_0)) - (\gamma - \beta \ctanh(\beta x_0))
 (\gamma + \beta \ctanh(\beta \bar{x}))}{q^2 + 2\beta \ctanh(\beta L)q + \beta^2 - \gamma^2}
\end{eqnarray*}
for $x_0 < x < L$.  Considering this expression as a function $q$, one
can decompose it into the sum of partial fractions and then perform
the inverse Laplace transform with respect to $q$ in order to get the
full propagator.  In this way, one can re-derive the general spectral
decomposition.

\begin{figure}
\begin{center}
\includegraphics[width=70mm]{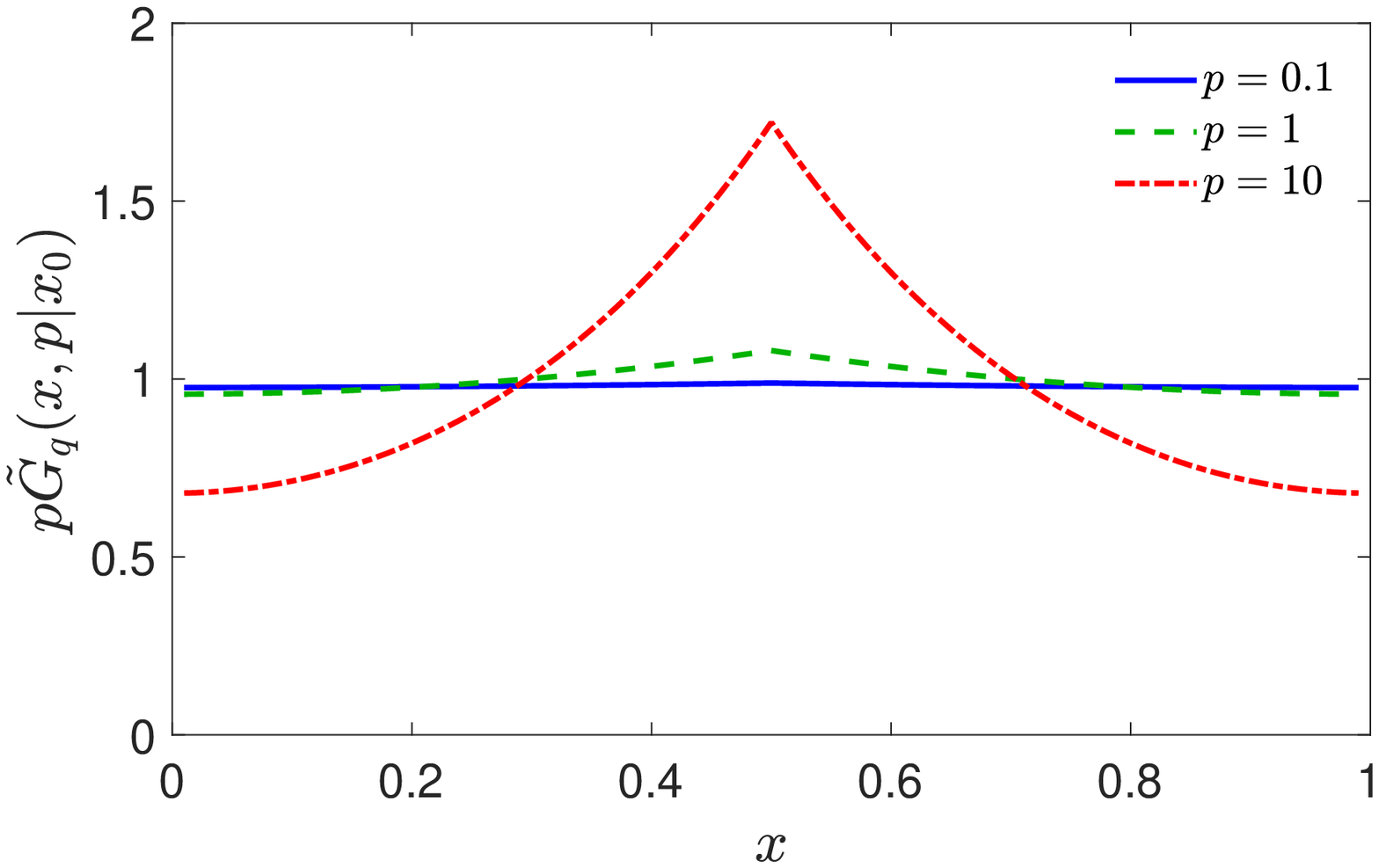} % Gq_q1e-3_mu0.eps}
\includegraphics[width=70mm]{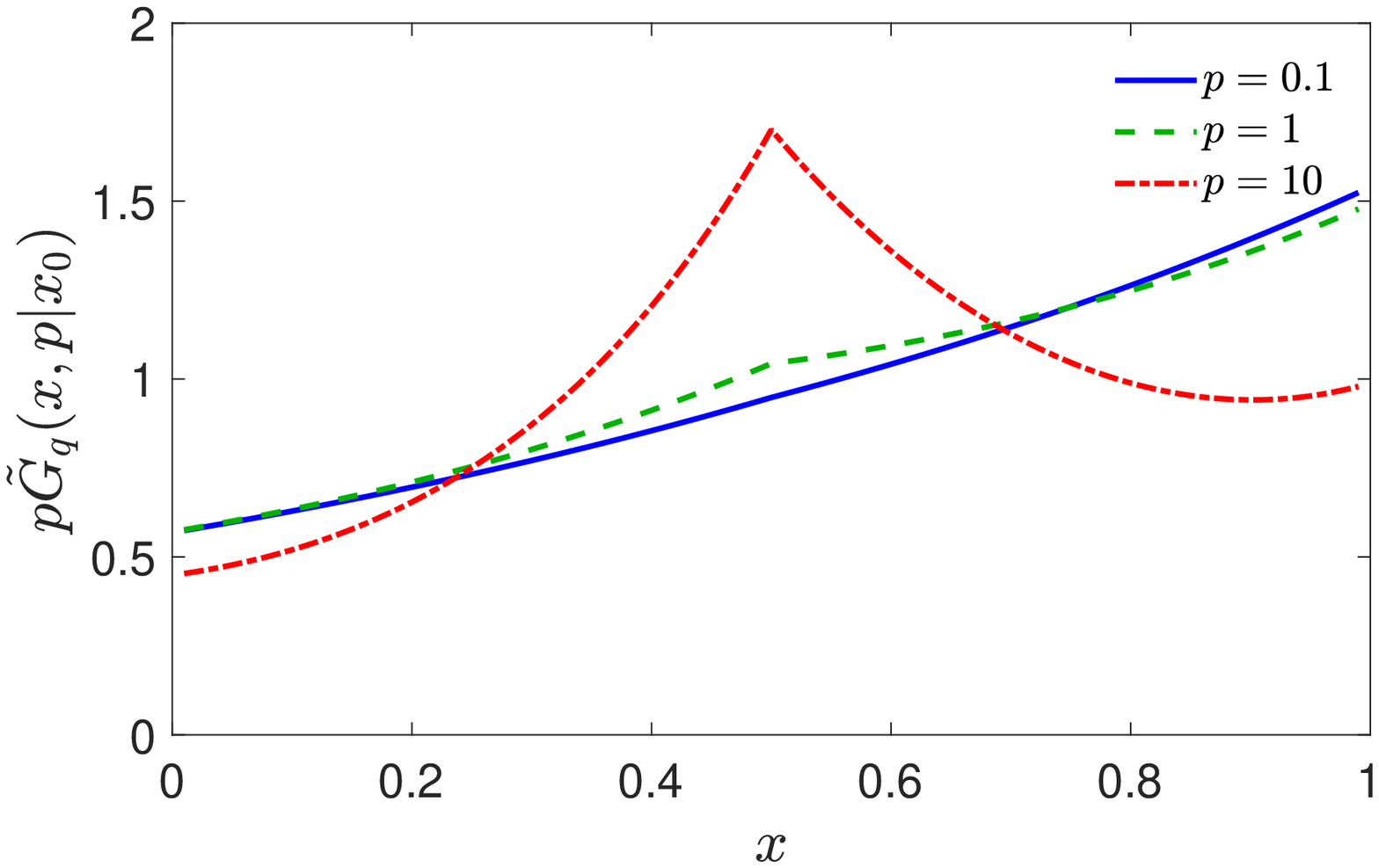} % Gq_q1e-3_mu1.eps}
\includegraphics[width=70mm]{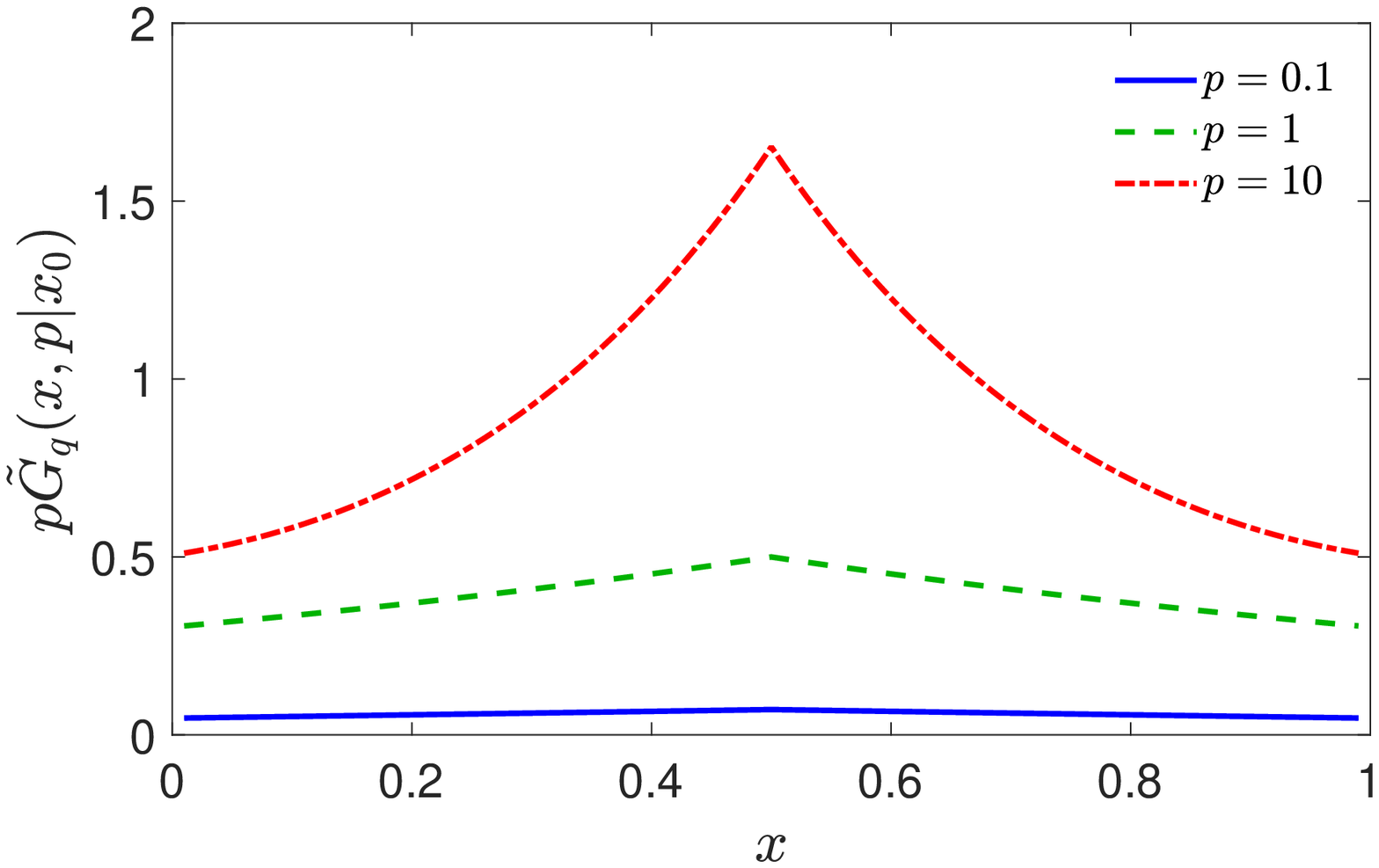} % Gq_q1_mu0.eps}
\includegraphics[width=70mm]{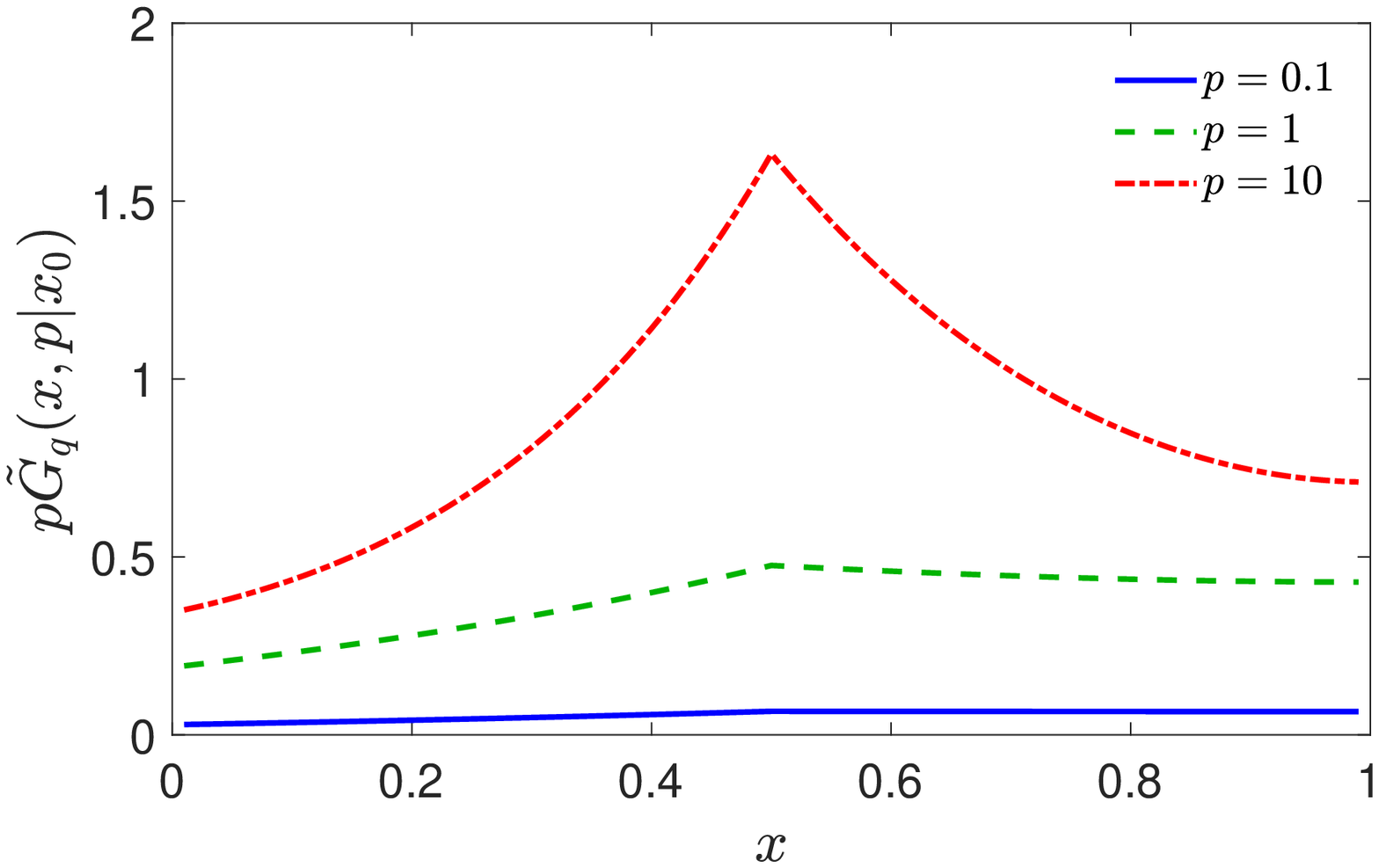} % Gq_q1_mu1.eps}
\includegraphics[width=70mm]{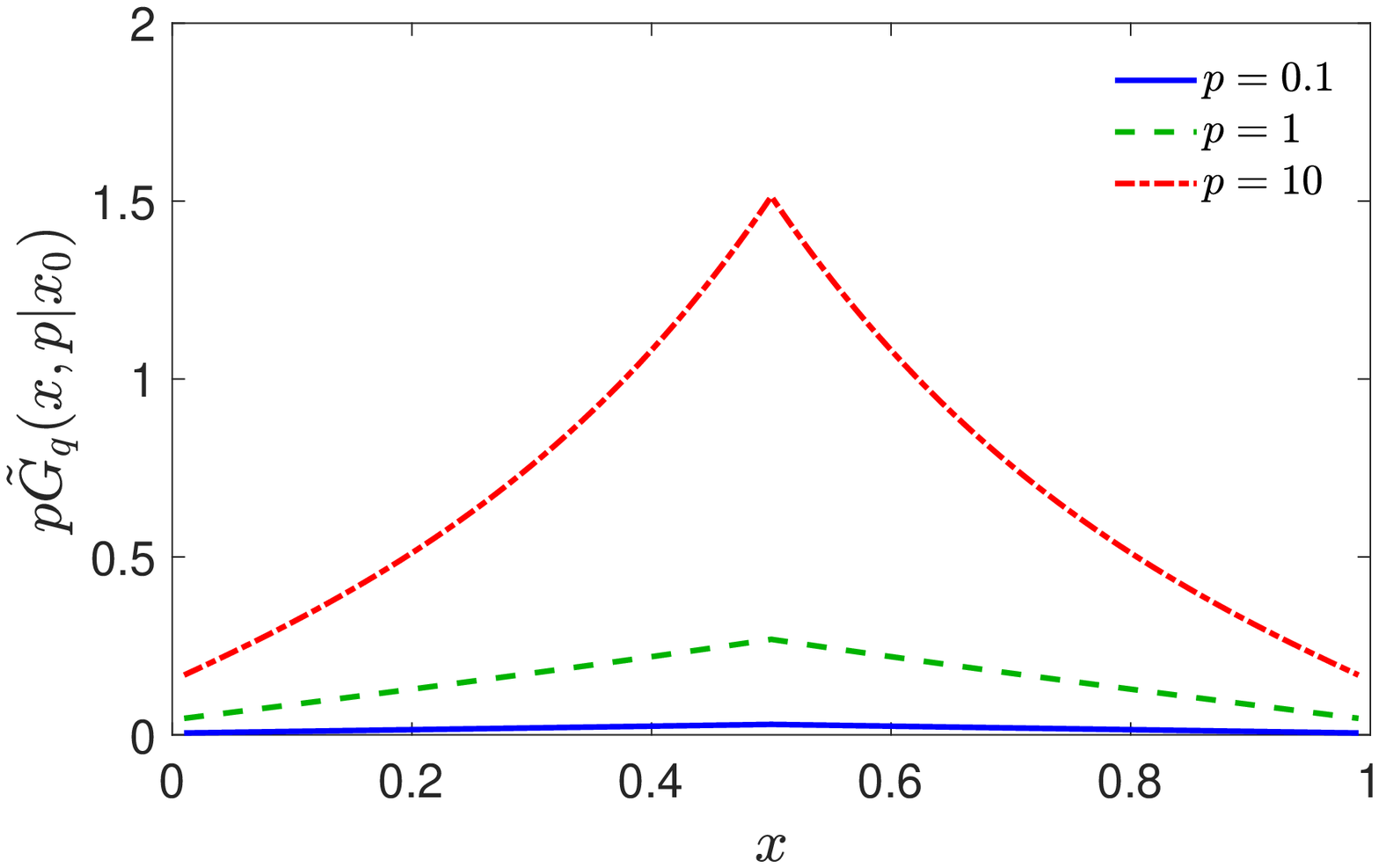} % Gq_q10_mu0.eps}
\includegraphics[width=70mm]{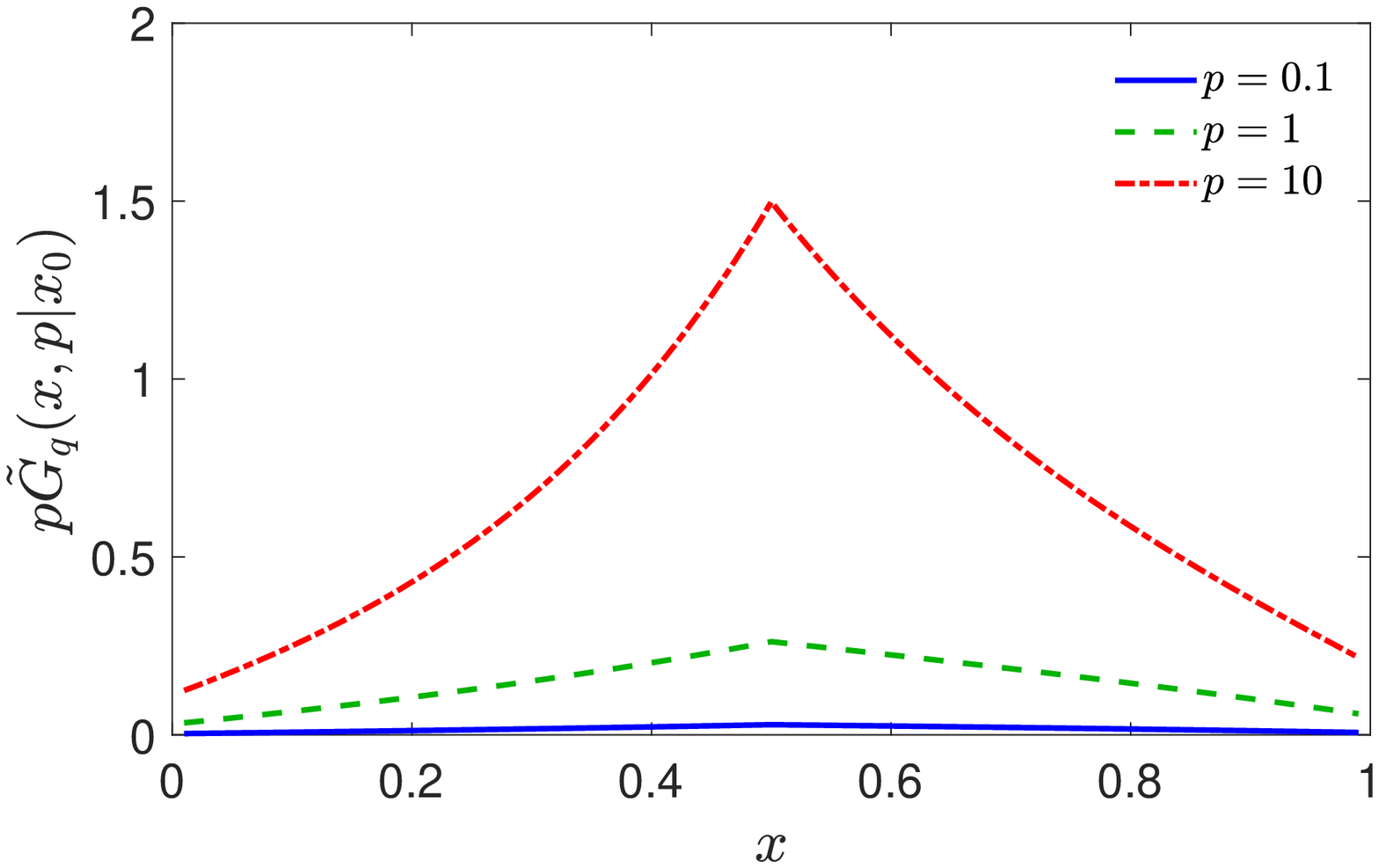} % Gq_q10_mu1.eps}
\end{center}
\caption{
Green's function $\tilde{G}_q(x,p|x_0)$ (multiplied by $p$) on the
unit interval ($L = 1$), with $D=1$, $x_0 = 0.5$, several values of
$p$.  Six panels correspond to $\mu = 0$ (left column) and $\mu = 1$
(right right) and $q = 10^{-3}$ (top row), $q = 1$ (middle row), and
$q = 10$ (bottom row). }
\label{fig:Gq}
% A_localtime8_drift_Gq_fig(1e-3,0);
% A_localtime8_drift_Gq_fig(1e-3,1);
\end{figure}

Figure \ref{fig:Gq} illustrates the behavior of the Green's function
$\tilde{G}_q(x,p|x_0)$ (for visual convenience, this function is
multiplied by $p$; in fact, for $q = 0$, the integral of this function
over $x$ is equal to $1/p$).  When $\mu = 1$, the right wing of
$\tilde{G}_q(x,p|x_0)$ is higher, as the positive drift facilitates
diffusion to the right.

\section{No drift limit}
\label{sec:no_drift}

In this Appendix, we briefly describe how the results of
Sec. \ref{sec:interval} are simplified in the no drift limit.

In the limit $\mu \to 0$, the eigenvalues of the Dirichlet-to-Neumann
operator tend to
\begin{equation}
\mu_{\pm}^{(p)} = \frac{\beta (\cosh(\beta L) \pm 1)}{\sinh(\beta L)}  \qquad \textrm{with} ~ \beta = \sqrt{p/D},
\end{equation}
while the eigenvectors tend to 
\begin{equation}
v_{\pm}^{(p)} = \frac{1}{\sqrt{2}} \left(\begin{array}{c} 1 \\ \mp 1 \\ \end{array} \right) ,
\end{equation}
so that
\begin{equation}
V_{\pm}^{(p)}(x) = V_{\pm}^{'(p)}(x) = \frac{\sinh(\beta \bar{x}) \mp \sinh(\beta x)}{\sqrt{2} \sinh(\beta L)} .
\end{equation}
In this way, we retrieve the Laplace-transformed full propagator for ordinary diffusion
without drift, first derived in \cite{Grebenkov20b}
\footnote{
There is a misprint in the expression (A.7) from \cite{Grebenkov20b}:
the sign minus in front of the second term should be replaced by plus,
as in Eq. (\ref{eq:Pfull_nodrift}).}
%footnote
%
\begin{eqnarray} \fl  \nonumber
&& \tilde{P}(x,\ell,p|x_0) = \tilde{G}_\infty(x,p|x_0) \delta(\ell) + \frac{e^{-\ell \beta \ctanh(\beta L)}}{D} \Biggl\{
\frac{\sinh(\beta \bar{x}_0) \sinh(\beta \bar{x}) + \sinh(\beta x_0) \sinh(\beta x)}{\sinh^2(\beta L)} \\  \fl   \label{eq:Pfull_nodrift}
&& \times \cosh\biggl(\frac{\beta \ell}{\sinh(\beta L)}\biggr)
+ \frac{\sinh(\beta \bar{x}_0) \sinh(\beta x) + \sinh(\beta x_0) \sinh(\beta \bar{x})}{\sinh^2(\beta L)} 
\sinh\biggl(\frac{\beta \ell}{\sinh(\beta L)}\biggr) \Biggr\}.
\end{eqnarray}
The marginal probability density of the boundary local time in the
Laplace domain reads
\begin{equation} \fl \label{eq:Pell_nodrift}
\tilde{P}(\circ,\ell,p|x_0) = \tilde{S}_\infty(p|x_0) \delta(\ell) 
+ \underbrace{\frac{\cosh(\beta L)-1}{\beta \sinh(\beta L)}}_{= \frac{\tanh(\beta L/2)}{\beta}} \, \frac{\sinh(\beta x_0)
+ \sinh(\beta \bar{x}_0)}{\sinh (\beta L)} \, \frac{e^{-\beta \tanh(\beta L/2) \ell}}{D} \,.
\end{equation}

\section{Asymptotic behavior}
\label{sec:short}

In this Appendix, we provide technical details for the asymptotic
analysis of the short-time and long-time regimes, which correspond to
the large-$p$ and small-$p$ limits in the Laplace domain,
respectively.

\subsection{Short-time behavior}

Under the condition $L\sqrt{p/D} \gg 1$, we get $\beta L \gg 1$ and
thus find $\mu_{\pm}^{(p)} \approx \beta \pm |\gamma|$, with
exponentially small corrections.  Similarly, we have $v_+^{(p)}
\approx (1,0)^\dagger$ and $v_-^{(p)} \approx (0,1)^\dagger$ for $\mu
> 0$, whereas the opposite relations hold for $\mu < 0$, namely,
$v_+^{(p)} \approx (0,1)^\dagger$ and $v_-^{(p)}
\approx (1,0)^\dagger$.  As a consequence, we get
\begin{equation}
V_+^{(p)}(x_0) \approx e^{\gamma x_0 - \beta x_0}  \,, \qquad
V_-^{(p)}(x_0) \approx e^{\gamma x_0 - \beta \bar{x}_0}   \qquad (\mu > 0),
\end{equation}
and
\begin{equation}
V_+^{(p)}(x_0) \approx e^{\gamma x_0 - \beta \bar{x}_0}  \,, \qquad
V_-^{(p)}(x_0) \approx e^{\gamma x_0 - \beta x_0}   \qquad (\mu < 0).
\end{equation}
Substituting these expressions into Eq. (\ref{eq:ellt_moments}), we get
\begin{equation} \label{eq:auxil67}
\int\limits_0^\infty dt\, e^{-pt} \, \E_{x_0}\{ \ell_t \} \approx
\frac{1}{p} \left(\frac{e^{-(\beta-\gamma)x_0}}{\beta - \gamma} 
+ \frac{e^{-(\beta+\gamma)(L-x_0)}}{\beta + \gamma}\right)  ,
\end{equation}
whatever the sign of $\mu$.
%\verb|A_localtime8_drift_Eell_p_fig;|]
%
Using the identity
\begin{equation}
\int\limits_0^\infty dt \, e^{-pt} \, e^{-a^2/(4t)} \left(\frac{1}{\sqrt{\pi t}} - b\, \erfcx\biggl(\frac{a}{2\sqrt{t}} + b\sqrt{t}\biggr)\right)
= \frac{e^{-a\sqrt{p}}}{\sqrt{p}+b} 
\end{equation}
(here $\erfcx(x) = e^{x^2} (1-\erf(x))$ is the scaled complementary
error function), the inverse Laplace transform of
Eq. (\ref{eq:auxil67}) reads
\begin{eqnarray*} \fl
\E_{x_0}\{ \ell_t \} &\approx& \int\limits_0^t dt' \,e^{-D\gamma^2 t'} \left\{ e^{\gamma x_0} \left[
e^{-x_0^2/(4Dt')} \left(\frac{\sqrt{D}}{\sqrt{\pi t'}} + \gamma D\, \erfcx\left(\frac{x_0}{\sqrt{4Dt'}} - \gamma\sqrt{Dt'}\right)\right)\right]
\right. \\ \fl
&+& \left. e^{\gamma(x_0-L)} \left[e^{-(L-x_0)^2/(4Dt')} \left(\frac{\sqrt{D}}{\sqrt{\pi t'}} 
- \gamma D\, \erfcx\left(\frac{L-x_0}{\sqrt{4Dt'}} + \gamma\sqrt{Dt'}\right)\right)\right] \right\} \\ \fl
&\approx& \int\limits_0^t dt' \, \left\{ e^{\gamma x_0} \left[
e^{-D\gamma^2 t'} e^{-x_0^2/(4Dt')} \frac{\sqrt{D}}{\sqrt{\pi t'}} + \gamma D\, e^{- \gamma x_0} 
\erfc\left(\frac{x_0}{\sqrt{4Dt'}} - \gamma\sqrt{Dt'}\right)\right] \right.  \\ \fl
&+& \left.  e^{-\gamma(L-x_0)} \left[e^{-D\gamma^2 t'} e^{-(L-x_0)^2/(4Dt')} \frac{\sqrt{D}}{\sqrt{\pi t'}} 
- \gamma D\, e^{\gamma(L-x_0)} \erfc\left(\frac{L-x_0}{\sqrt{4Dt'}} + \gamma\sqrt{Dt'}\right)\right] \right\} .
\end{eqnarray*}
Using the identity
\begin{equation*} \fl
\int\limits_0^t dt' \frac{e^{-at'-b/t'}}{\sqrt{t'}} = \frac{\sqrt{\pi}}{2\sqrt{a}} \biggl(e^{2\sqrt{ab}} \biggl(\erf(\sqrt{at}+\sqrt{b/t}) - 1\biggr)
+ e^{-2\sqrt{ab}} \biggl(\erf(\sqrt{at}-\sqrt{b/t}) + 1\biggr)\biggr),
\end{equation*}
we get
\begin{eqnarray*} \fl
\E_{x_0}\{ \ell_t \} &\approx&  
\frac{e^{\gamma x_0}}{2|\gamma|} \biggl[- e^{|\gamma|x_0} \erfc(x_0/\sqrt{4Dt}+|\gamma|\sqrt{Dt}) 
+ e^{-|\gamma|x_0} \erfc(x_0/\sqrt{4Dt} - |\gamma|\sqrt{Dt}) \biggr] \\ \fl
&+& \frac{e^{-\gamma \bar{x}_0}}{2|\gamma|} \biggl[- e^{|\gamma|\bar{x}_0} \erfc(\bar{x}_0/\sqrt{4Dt}+|\gamma|\sqrt{Dt}) 
+ e^{-|\gamma|\bar{x}_0} \erfc(\bar{x}_0/\sqrt{4Dt} - |\gamma|\sqrt{Dt})\biggr] \\ \fl
&+& \gamma D \int\limits_0^t dt' \biggl\{ \erfc\biggl(\frac{x_0}{\sqrt{4Dt'}} - \gamma \sqrt{Dt'}\biggr)
- \erfc\biggl(\frac{\bar{x}_0}{\sqrt{4Dt'}} + \gamma\sqrt{Dt'}\biggr) \biggr\} .
\end{eqnarray*}
In particular, for $x_0 = 0$, we get after simplifications
\begin{equation}  \label{eq:Eell0_short}
\E_{0}\{ \ell_t \} \approx \frac{1}{\gamma} \left( \frac{\erf(\gamma\sqrt{Dt})}{2} + \gamma^2 Dt \,\erfc(-\gamma\sqrt{Dt}) 
- \frac{|\gamma| \sqrt{Dt}}{\sqrt{\pi}}\, e^{-D\gamma^2 t}\right).
\end{equation}

\subsection{Long-time behavior}

Let us consider the case $\gamma \ne 0$.  In the limit $p\to 0$, we
get up to $O(p^2)$:
\begin{eqnarray*}
\beta &\approx& |\gamma| + \frac{p}{2|\gamma|D} \,, \\
%\mu_{\pm}^{(p)} &\approx& |\gamma|\ctanh(|\gamma|L) \biggl(1 + \frac{p}{2\gamma^2D} - \frac{pL}{|\gamma|D \sinh(2|\gamma|L)}\biggr) \\
%&\pm& \biggl(\gamma^2 \ctanh^2(|\gamma|L) + \frac{\gamma^2}{\sinh^2(|\gamma|L)} \biggl(\frac{p}{D\gamma^2} - \frac{pL}{|\gamma|D} \ctanh(|\gamma|L)\biggr)
%\biggr)^{1/2} \,, \\
\mu_{\pm}^{(p)} &\approx& |\gamma|\ctanh(|\gamma|L) \biggl(1 + \frac{p}{2\gamma^2D} - \frac{pL}{|\gamma|D \sinh(2|\gamma|L)}\biggr) \\
&\pm& |\gamma| \ctanh(|\gamma|L)\biggl(1 + \frac{1}{2\cosh^2(|\gamma|L)} \biggl(\frac{p}{D\gamma^2} - \frac{pL}{|\gamma|D} \ctanh(|\gamma|L)\biggr)
\biggr) \,,
\end{eqnarray*}
so that
\begin{eqnarray*}
\mu_-^{(p)} &\approx& p \frac{\tanh(\gamma L)}{2\gamma D} + O(p^2),  \\
\mu_+^{(p)} &\approx& 2\gamma \ctanh(\gamma L) + O(p).
\end{eqnarray*}
We also have
\begin{eqnarray*}
(\M_p)_{11} &\approx& \gamma \ctanh(\gamma L) \left(1 + \frac{p}{2\gamma^2D} - \frac{pL}{\gamma D \sinh(2 \gamma L)}\right) - \gamma , \\
(\M_p)_{12} &\approx& - \frac{\gamma}{\sinh(\gamma L)} \left(1 + \frac{p}{2\gamma^2D} - \frac{pL \ctanh(\gamma L)}{2\gamma D}\right).
\end{eqnarray*}
In the lowest order, we also get
\begin{eqnarray*}
v_-^{(p)}(0) &\approx& \frac{1}{\cosh(\gamma L) \sqrt{2(1 - \tanh(\gamma L))}} + O(p),\\
v_-^{(p)}(L) &\approx& \sqrt{(1 - \tanh(\gamma L))/2} + O(p) .
\end{eqnarray*}
Substituting these expressions into Eq. (\ref{eq:Eell}) for $x_0$, we
deduce in the leading order 
\begin{equation}
\int\limits_0^\infty dt \, e^{-pt} \, \E_{0}\{ \ell_t\} = \frac{2\gamma D \ctanh(\gamma L)}{p^2} + O(1/p),
\end{equation}
from which Eq. (\ref{eq:Eell_t_long}) follows immediately.

\section{On the numerical inversion of the Laplace transform}
\label{sec:Talbot}

The inverse Laplace transform $f(t)$ of a function $\tilde{f}(p)$ can
be expressed as the Bromwich integral,
\begin{equation}
f(t) = \frac{1}{2\pi i} \int\limits_{\gamma} dp \, e^{pt} \, \tilde{f}(p),
\end{equation}
where $\gamma$ is a contour from $-i\infty$ to $+i\infty$ in the
complex plane that lies on the right to singularities of
$\tilde{f}(p)$.  In the basic Talbot algorithm, the contour $\gamma$
is parameterized as 
\begin{equation}
\gamma~:~ \theta \to p(\theta) = \sigma + \mu(\theta \ctan(\theta) + \nu i \theta),  \qquad \theta \in (-\pi,\pi),
\end{equation}
with three parameters $\sigma$, $\mu$ and $\nu$ whose optimal choice
depends on the function $\tilde{f}(p)$ and determines the convergence
rate \cite{Talbot79}.  Using this contour, the Bromwich integral can
be reduced to
\begin{equation}
f(t) = \frac{1}{2\pi i} \int\limits_{-\pi}^{\pi} d\theta \, \frac{dp}{d\theta} \, e^{p(\theta) t} \, \tilde{f}(p(\theta)),
\end{equation}
and then numerically evaluated by quadratures.  In this paper, we used
a slightly different contour by Weideman \cite{Weideman06}:
\begin{equation}  \fl
\gamma~:~ \theta \to p(\theta) = N \bigl(0.5017\, \theta \,\ctan(0.6407\, \theta) -0.6122 + 0.2645 i\, \theta),  \quad \theta \in (-\pi,\pi),
\end{equation}
where the value of $N$ controls the accuracy of computation.

%%%  See ILT_Weideman3(psip, t, Nt);

\vskip 10mm
%\newpage

\end{document}